\documentclass{emulateapj}

\shorttitle{TEXES SURVEY FOR H$_{2}$ EMISSION FROM DISKS} 
\shortauthors{BITNER ET AL.}

\newcommand{\szero}{S(0)}
\newcommand{\sone}{S(1)}
\newcommand{\stwo}{S(2)}
\newcommand{\sthree}{S(3)}
\newcommand{\sfour}{S(4)}
\newcommand{\seight}{S(8)}
\newcommand{\snine}{S(9)}
\newcommand{\escm}{ergs s$^{-1}$ cm$^{-2}$}

\begin{document}

\title{THE TEXES SURVEY FOR H$_{2}$ EMISSION FROM PROTOPLANETARY DISKS}

\author{Martin A. Bitner\altaffilmark{1,2}, Matthew J. Richter\altaffilmark{2,3}, 
John H. Lacy\altaffilmark{1,2}, Gregory J. Herczeg\altaffilmark{4}, 
Thomas K. Greathouse\altaffilmark{2,5}, Daniel T. Jaffe\altaffilmark{1,2}, 
Colette Salyk\altaffilmark{6}, Geoffrey A. Blake\altaffilmark{6}, 
David J. Hollenbach\altaffilmark{7}, Greg W. Doppmann\altaffilmark{8}, 
Joan R. Najita\altaffilmark{8}, Thayne Currie\altaffilmark{9}}

\altaffiltext{1}{Space Telescope Science Institute, Baltimore, MD; 
mbitner@stsci.edu}
\altaffiltext{2}{Visiting Astronomer at the Infrared Telescope Facility, which is operated by the University of Hawaii under Cooperative Agreement no. NCC 5-538 with the National Aeronautics and Space Administration, Science Mission Directorate, Planetary Astronomy Program.}
\altaffiltext{3}{Physics Department, University of California at Davis, Davis, CA }
\altaffiltext{4}{California Institute of Technology, Pasadena, CA}
\altaffiltext{5}{Southwest Research Institute, San Antonio, TX}
\altaffiltext{6}{Division of Geological \& Planetary Sciences, California Institute 
of Technology, MS 150-21, Pasadena, CA}
\altaffiltext{7}{NASA Ames Research Center, Moffett Field, CA}
\altaffiltext{8}{National Optical Astronomy Observatory, Tucson, AZ}
\altaffiltext{9}{Harvard-Smithsonian Center for Astrophysics, Cambridge, MA}

\begin{abstract}

We report the results of a search for pure rotational molecular hydrogen emission 
from the circumstellar environments of young stellar objects with disks using the Texas 
Echelon Cross Echelle Spectrograph (TEXES) on the NASA Infrared Telescope 
Facility and the Gemini North Observatory.
We searched for mid-infrared H$_{2}$ emission in the \sone, \stwo, and \sfour ~transitions.
Keck/NIRSPEC observations of the H$_{2}$ \snine ~transition were included for some sources 
as an additional constraint on the gas temperature.  
We detected H$_{2}$ emission from 6 of 29 sources observed: AB Aur, DoAr 21, 
Elias 29, GSS 30 IRS 1, GV Tau N, and HL Tau.
Four of the six targets with detected emission are class I sources that show 
evidence for surrounding material in an envelope in addition to a circumstellar disk.
In these cases, we show that accretion shock heating is a plausible excitation mechanism.
The detected emission lines are narrow ($\sim$10 km s$^{-1}$), 
centered at the stellar velocity, and spatially unresolved at scales of 0.4\arcsec, which is 
consistent with origin from a disk at radii 10-50 AU from the star.
In cases where we detect multiple emission lines, we derive temperatures $\gtrsim$ 500 K 
from $\sim$1 M$_{\oplus}$ of gas.
Our upper limits for the non-detections place upper limits on the amount of
H$_2$ gas with T $>$500~K of less than a few Earth masses.
Such warm gas temperatures are significantly higher than the equilibrium dust temperatures 
at these radii, suggesting that the gas is decoupled from the dust in the regions 
we are studying and that processes such as UV, X-ray, and accretion heating may be important.
 
\end{abstract}



%++++++++++++++++++++++++++++++++++++++++++++++++++++++++++
%  INTRODUCTION
%++++++++++++++++++++++++++++++++++++++++++++++++++++++++++

\section{INTRODUCTION}

Studying the structure and evolution of circumstellar disks is crucial to 
developing an understanding of the process of planet formation.
Observations of dust emission and modeling of the spectral energy 
distributions (SED) of disks have revealed much about the dust component from 
a few stellar radii out to hundreds of AU \citep{zuckerman01}.
While circumstellar disks are composed of both dust and gas, the gas component 
dominates the mass of the disk, with molecular hydrogen (H$_{2}$) being 
the most abundant constituent.
In order to develop a complete picture of the structure and evolution 
of protoplanetary disks, it is important to observe the gas.

Observations at different wavelengths probe different disk radii.
Submillimeter observations sample gas at large radii ($> 50$ AU) 
\citep{semenov05}, while near-infrared CO \citep{najita03,blake04} and H$_{2}$O
\citep{carr04, thi05} observations allow for study of the inner few AU.
Spectral lines in the mid-infrared (5-25 $\mu$m) provide a means to investigate gas in the 
giant planet region of the disk and beyond (10-50 AU) \citep{najita07a}.

Several mid-infrared spectral diagnostics have been shown to be
useful probes of gas in disks.  These include 
[NeII] at 12.8 $\mu$m \citep{pascucci07,lahuis07,herczeg07}, H$_{2}$O 
rotational transitions \citep{carr08,salyk08}, [FeI] at 
24 $\mu$m \citep{lahuis07}, and, based on a theoretical analysis of 
debris disks, [SI] at 25.2 
$\mu$m, and [FeII] at 26 $\mu$m \citep{gorti04}.

Molecular hydrogen should make up the bulk of the mass in disks, 
but is a challenge to detect.
Bright far-ultraviolet (FUV) H$_{2}$ emission from classical T Tauri stars may 
be produced in the irradiated disk surface \citep{herczeg02,bergin04}.
At longer wavelengths, 
rovibrational and pure rotational transitions are generally weak
because H$_2$ lacks a permanent dipole moment.
Near-infrared emission in the \textit{v}=1-0 \sone ~rovibrational transition of H$_{2}$ 
has been detected from T Tauri stars \citep{bary03,ramsay07,carmona08b} and may be the 
result of excitation by UV and X-ray irradiation \citep{nomura07,gorti08}.
Near-infrared adaptive optics fed, integral field spectroscopy of six classical T Tauri 
stars that drive powerful outflows has revealed that most of the H$_{2}$ emission is 
spatially extended from the continuum \citep{beck08}.
The properties of the emission are consistent with shock excitation from 
outflows or winds rather than UV or X-ray excitation from the central star.
The FUV H$_{2}$ emission probes gas between 2000 and 3000 K \citep{herczeg04,nomura05} 
and the 2.12 $\mu$m H$_{2}$ line traces gas at T$>$1000 K \citep{bary03}.
The mid-infrared H$_{2}$ lines considered in this paper are most sensitive to gas at 
lower temperatures.

Owing to their small Einstein A-values, the pure rotational mid-infrared 
H$_{2}$ lines remain optically thin to large column densities 
(N$_{H_{2}} > 10^{23}$ cm$^{-2}$) and will be in LTE at the densities found in 
disks.
For a disk with a strong mid-infrared continuum, the dust becomes 
optically thick well before the H$_{2}$ lines.  
Observable line emission is present only when there is a temperature inversion 
in the atmosphere of the disk or if there is a layer of dust-depleted gas separate from 
the optically thick dust.
The process of dust coagulating into larger grains or 
settling out of the disk atmosphere can allow a larger 
column of gas to be observed.
A disk that has an optically thin mid-infrared continuum, implying a very small amount of
dust in the disk or dust grains that have grown large compared
to mid-infrared wavelengths, would allow the entire disk to be observable.
However, it is not known whether such disks have large quantities of gas and, 
in the absence of gas heating through collisions with dust grains, 
another heating mechanism is necessary in dust-free environments, such as UV or X-ray 
heating \citep{glassgold04,gorti04,nomura07}.

A number of groups have searched for H$_{2}$ emission from protoplanetary disks 
in recent years.  
\citet{thi01} reported the detection of several Jupiter masses of warm gas in a sample 
of disk sources based on \textit{Infrared Space Observatory} (ISO) observations 
of the H$_{2}$ \szero ~and \sone ~lines.
However, follow-up observations from the ground with 
improved spatial resolution did not confirm these results \citep{richter02,sheret03,sako05}.  
\citet{richter02} used the Texas Echelon Cross Echelle Spectrograph 
(TEXES) on the NASA 3m Infrared Telescope Facility 
(IRTF) to set upper limits on the warm gas mass within the disks of six young 
stars.
\citet{sheret03} searched for H$_{2}$ emission using MICHELLE on the United 
Kingdom Infrared Telescope and set upper limits on the emission 
from disks around two stars.  
The first group to use an 8-meter class telescope in the search for molecular 
hydrogen was \citet{sako05} using the Cooled Mid-Infrared Camera and 
Spectrometer on the 8.2 m Subaru telescope to set upper limits 
for emission in the \sone ~line around four young stars.
Using the Infrared Spectrograph (IRS) aboard the \textit{Spitzer Space Telescope}, 
\citet{pascucci06} reported the non-detection of H$_{2}$ lines in their sample 
of 15 young Sun-like stars, while \citet{lahuis07} detected the \stwo ~and 
\sthree ~lines in $\sim$8\% of the 76 circumstellar disks in their sample.
Recently, ground-based observations with high resolution spectrometers on 8-meter class 
telescopes have produced both detections of 
the mid-infrared H$_{2}$ lines in the Herbig Ae stars AB Aur and HD 97048 
\citep{bitner07,martin07} and stringent upper limits in a sample of six Herbig Ae/Be stars 
and one T Tauri star \citep{carmona08a}.  

Three mid-infrared pure rotational H$_{2}$ lines are observable from the ground:  
\sone ~($\lambda = 17.035 ~\mu$m), \stwo ~($\lambda = 12.279 ~\mu$m), and
\sfour ~($\lambda = 8.025 ~\mu$m).  
When multiple optically thin lines are observed, line ratios permit the determination of the 
temperature and mass of the emitting gas.
Ratios of these three lines are most sensitive to temperatures of 200-800 K.
Two additional pure rotational H$_{2}$ lines are observable near 5 $\mu$m:  
\seight ~($\lambda = 5.053 ~\mu$m) and \snine ~($\lambda = 4.695 ~\mu$m), extending 
our temperature sensitivity to hotter gas.
The high spectral resolution possible with an instrument like 
TEXES (Lacy et al. 2002) increases our sensitivity to narrow line emission and helps 
determine the location of the emission.
By making observations at high spectral resolution, we maximize the line to 
continuum contrast while minimizing atmospheric effects by separating 
the lines from nearby telluric features.
A further benefit of high spectral resolution is that we are able to estimate 
the location of the emitting gas if coming from a disk under the 
assumption of Keplerian rotation.

In this paper, we present the results of a search for molecular hydrogen emission 
in disk sources using TEXES on both the IRTF and Gemini North telescopes.
We observed 29 sources spanning a range of mass, age, and accretion rate in order 
to constrain the amount of warm gas in the circumstellar disks of these stars.

%++++++++++++++++++++++++++++++++++++++++++++++++++
%  OBSERVATIONS AND DATA REDUCTION
%++++++++++++++++++++++++++++++++++++++++++++++++++

\section{OBSERVATIONS AND DATA REDUCTION}

We observed 29 sources using TEXES on both the NASA IRTF telescope during 
2002-2005 and on Gemini North in 2006 and 2007 under program IDs GN-2006A-DS-3, GN-2006B-Q-42, 
and GN-2007B-C-9.
The spectral resolution of the observations was $\gtrsim$60,000 for the \sone ~line and 
$\gtrsim$80,000 for the \stwo ~and \sfour ~lines.
Due to telluric absorption from water vapor close to the \sone ~and \stwo ~lines, we observed 
these settings both when telluric water vapor levels were low and the Earth's motion 
gave an additional redshift towards the source.
Removal of background sky emission was achieved by nodding the source 
along the slit and subtracting nod pairs.
On the IRTF, the TEXES slit widths were 2\arcsec ~at the \sone ~setting and 1.4\arcsec ~at \stwo ~and \sfour.
For our Gemini observations, the slit widths were 0.81\arcsec ~at \sone ~and 0.54\arcsec ~at \stwo ~and \sfour.

We observed telluric standards at each setting for use as divisors to correct 
the spectra for atmospheric absorption.
Asteroids work well as telluric calibrators at 12 and 17 $\mu$m, while we used 
early-type stars at 8 $\mu$m.
We observed flux standards for some sources.
Where possible, however, we normalized the continuum level to agree with photometric 
measurements obtained from \textit{Spitzer}, \textit{ISO}, or ground-based observations.
Tables~\ref{tab:lines},~\ref{tab:ulirtf}, and~\ref{tab:ulgemini} list references for the continuum fluxes.
Data reduction was carried out using the standard TEXES pipeline \citep{lacy02}. 

Our sample was chosen to include sources with a range in age, accretion rate, and mass.
Included among our targets are class I sources with both a remnant envelope and a disk, 
classical T Tauri stars with optically thick 
disks, T Tauri stars with optically thin disks or inner holes, stars with high accretion 
rates such as FU Ori and Z CMa, Herbig Ae stars with more massive disks, as well as the 
debris disk around the star 49 Ceti.
The combination of TEXES with Gemini has allowed us to extend our survey to 
sources with mid-infrared continuum levels of a few tenths of a Jansky. 
Since the mid-infrared continuum comes from warm dust grains, this allows the 
inclusion of sources without much dust where larger columns of gas may be 
observable as long as there are additional heating mechanisms beyond gas/dust heating.
Properties of the targets in our sample are listed in Table~\ref{tab:props}.

Observations of the H$_{2}$ \snine ~line for a subset of our sample were carried 
out between 2000 and 2005 using NIRSPEC on the Keck telescope as part of an ongoing 
M-band survey \citep{blake04} at a spectral resolution of R=25,000.
The \snine ~observations were carried out using a slit width of 0.43\arcsec.
For a description of the data reduction process for these observations see \citet{boogert02b}.

%++++++++++++++++++++++++++++++++++++++++++++++++++
%  RESULTS
%++++++++++++++++++++++++++++++++++++++++++++++++++
\section{RESULTS}
Table~\ref{tab:lines} summarizes our line detections.
We detected H$_{2}$ emission from 6 of 29 sources observed: AB Aur, DoAr 21, 
Elias 29, GSS 30 IRS 1, GV Tau N, and HL Tau.
Fluxes, line widths, and centroids were determined by fitting each line individually 
with a Gaussian profile, which describes the detected lines reasonably well
in most cases.
In four of the five sources observed from both the IRTF and Gemini, the line fluxes 
or upper limits from Gemini are smaller than those measured from the IRTF.
Since the TEXES slit is larger on the sky at the IRTF, this suggests the H$_{2}$ 
emission may be spatially extended.
Figures~\ref{gem-abaur-s9s4s2s1}-\ref{gem-hltau-s9s4s2s1} show that the lines are all centered 
near the stellar velocity suggesting the H$_{2}$ is associated with the targets.
In all cases, the mid-infrared lines are narrow with observed FWHM near 10 km s$^{-1}$.
Correction for the TEXES instrumental line width of 4-5 km s$^{-1}$ would give source 
line widths $\sim$1 km s$^{-1}$ smaller than observed.
For the Keck/NIRSPEC \snine ~observations listed in Table~\ref{tab:lines}, 
removal of the instrumental line broadening would give line widths 3-4 km s$^{-1}$ 
narrower than observed.
Although our spectra 
cover $\pm$100~km~s$^{-1}$ from the stellar velocity
for the \stwo ~line, and a similar amount to the red of 
the \sone ~line, we see no evidence for emission at large Doppler shifts. 
We most often obtain the highest signal-to-noise detections of the \stwo ~line, where the
atmospheric transmission is higher and TEXES is more sensitive.  

For emission from optically thin gas, the line flux is given by 
\begin{equation}
   {F= \frac{N_{u} \times A_{ul} \times h \nu} {4 \pi \times d^{2}}},
        \label{flux-eq}
\end{equation}
where $N_{u}$ is the number of 
H$_{2}$ molecules in the upper state, $\nu$ is the line frequency, and $d$ 
is the distance to the source.
The gas mass and temperature enter the equation determining the number of molecules 
in the upper state, which, under the assumption of LTE at a single temperature, takes the form
\begin{equation}
  {N_{u} = \frac{M}{m_{H_{2}}} \times 
  g_{N}(2J_{u}+1) \frac{hcB}{2kT} e^{\frac{-E_{u}}{kT}}},
        \label{Nu-eq}
\end{equation}
where $M$ is the H$_{2}$ gas mass, $m_{H_{2}}$ is the mass of an 
H$_{2}$ molecule, $g_{N}$ is the nuclear 
statistical weight (3 or 1 for ortho or para), $B$ is the rotational constant for H$_{2}$ 
taken from \citet{jennings84}, and $E_{u}$ is the upper state energy taken from \citet{mandy93}.

\begin{figure}
\plotone{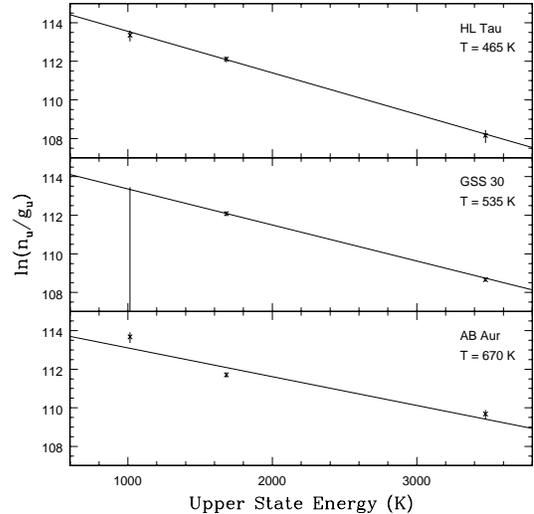}
%\centerline{f8.eps}
\figcaption[f8.eps]
{Excitation diagrams for the three sources in our sample where we have 
observations of all three mid-infrared H$_{2}$ transitions and detections of at least two.
The points are based on Gaussian fits to each of the transitions and are plotted with 
1-$\sigma$ error bars.  The overplotted lines show the best-fit single temperature.
\label{popdia}
}
\end{figure}

In Table~\ref{tab:ltegemini} we show the derived temperature and mass for the emitting gas 
in each source based on both single temperature and two-temperature LTE models.
Figure~\ref{popdia} shows excitation diagrams for the three sources in our sample where we have 
observations of all three mid-infrared H$_{2}$ transitions and detections of at least two.
We assumed optically thin LTE H$_{2}$ emission and constructed synthetic spectra for a range 
of temperatures and masses using the Gaussian line parameters determined from fits to 
each line individually.
We simultaneously fit the synthetic spectra to all of the data for each source, including 
non-detections, and determined the best fit by minimizing the square of the residuals.
The errors listed in Table~\ref{tab:ltegemini} are 1$\sigma$ based on the contour plot 
of the $\chi^{2}$ values.
We present the results of our best fitting model in the appropriate
figures, but defer discussion of the individual sources until the next section.

Tables~\ref{tab:ulirtf} and \ref{tab:ulgemini} list our derived 3$\sigma$ 
line flux upper limits for our IRTF and Gemini observations in the cases 
where no line was detected.  
The standard deviation ($\sigma$) of the line fluxes was computed by looking at the distribution 
of values found when assuming a FWHM comparable to our line detections (10~km~s$^{-1}$) 
and summing over the number of pixels corresponding to that FWHM in 
regions of the spectrum with comparable atmospheric transmission.

\begin{figure}
\plotone{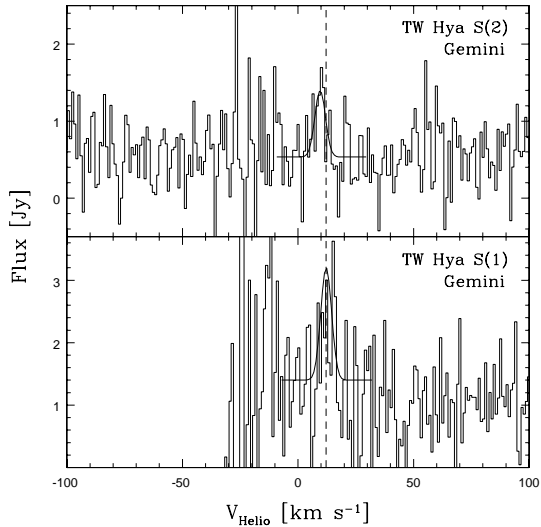}
%\centerline{f9.eps}
\figcaption[f9.eps]
{TW Hya TEXES/Gemini S(2) and S(1) overplotted with a Gaussian fit to the 
(2$\sigma$) bump near the S(2) position and the 3$\sigma$ upper limit at S(1).
The measured FWHM of 5.5 km s$^{-1}$ for the Gaussian fit at S(2) was assumed 
to obtain the line flux upper limit at S(1).
The S(1) and S(2) upper limits are listed in Table~\ref{tab:ulgemini}.
The dashed line indicates the systemic velocity of TW Hya \citep{kastner99}. 
\label{twhya-s2s1}
}
\end{figure}

In the case of TW Hya, we used a FWHM of 5.5 km s$^{-1}$ to compute the line 
flux upper limits based on the measured width of the (2$\sigma$) \stwo ~feature.
Figure~\ref{twhya-s2s1} shows observations of TW Hya made during the TEXES/Gemini 
engineering run in February 2006.
There is a hint of a feature near the \stwo ~position but not a clear detection.
Figures~\ref{irtf-s1},~\ref{irtf-s2},~\ref{gem-s1},~and \ref{gem-s2} 
show 3$\sigma$ \sone ~and \stwo ~line flux upper limits assuming a 
FWHM for the lines of 10 km s$^{-1}$ overplotted on the 
continuum at the stellar velocity.
The gas mass upper limits listed in Tables~\ref{tab:ulirtf} and ~\ref{tab:ulgemini} were 
computed under the assumption that the H$_{2}$ is in local thermodynamic equilibrium (LTE) 
and that the emission is optically thin.

At the temperatures to which we are sensitive, T $>$ 200 K, the LTE value of the ortho/para 
ratio (OPR) is 3, which we assume for our calculations.
However, nonequilibrium values of the OPR have been observed in gas 
at these temperatures.
Using \textit{ISO}, \citet{neufeld98} derived an OPR of 1.2 towards HH 54 
from gas at T $\sim$ 650 K.
They argued that the warm, shocked H$_{2}$ gas they observed acquired its OPR
 at T $<$ 90 K and that it has not had time to equilibrate to the LTE value 
at the higher temperature.
\citet{fuente99} found an OPR between 1.5 and 2 from 300-700 K gas 
in a photodissociation region (PDR), which is of interest since the surface layers 
of the disks we may be observing have similarities with PDRs \citep{jonkheid04}.
\citet{bitner07} claimed that the surface brightness of the H$_{2}$ emission 
observed towards the disk source AB Aur was similar to the Orion bar PDR, lending 
support to the idea that PDRs and disk surfaces have similar qualities.
This was the result of a computation error.  
In fact, the surface brightness of the H$_{2}$ emission from the Orion bar PDR is 
significantly larger than that of the AB Aur emission.
We note that FUV pumping can lead to an apparent OPR in this range even in gas with 
an equilibrium OPR of 3 due to ortho-H$_{2}$ pumping rates being reduced by 
self-shielding \citep{sternberg99}.
If the OPR is actually less than 3 in the sources in our sample, a single-temperature 
fit to our observations would lead us to derive a higher gas temperature 
because the lowest energy transition we observe is the ortho \sone.

As expected, the upper limits are generally more stringent for the Gemini data than for the IRTF data.
The derived 3$\sigma$ line flux upper limits for our Gemini data are in the range 
of 10$^{-15}$ to 10$^{-14}$ ergs s$^{-1}$ cm$^{-2}$.
Observations with IRS on \textit{Spitzer} of sources with low continuum fluxes 
give 3$\sigma$ line flux upper limits of 10$^{-16}$ to 10$^{-14}$ \escm ~\citep{pascucci06,lahuis07}.
The very low upper limits possible with \textit{Spitzer} are for sources with 
continuum fluxes too low to be detected from the ground at the high
spectral resolution available with TEXES.
In one case where we have a source in common, 49 Cet, our derived upper limits 
are nearly identical to those obtained with \textit{Spitzer} \citep{chen06}.  
Our upper limits constrain the amount of warm gas with optically thin dust to be below 
several tens of Earth masses for a temperature of 200 K and less than a few Earth 
masses at temperatures above 500 K.

%++++++++++++++++++++++++++++++++++++++++++++++++++
%  DISCUSSION
%++++++++++++++++++++++++++++++++++++++++++++++++++

\section{DISCUSSION}

\begin{figure}
\plotone{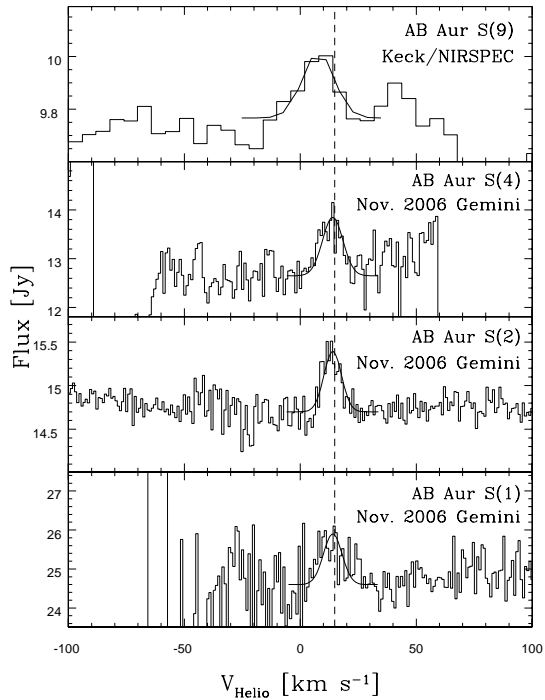}
%\centerline{f1.eps}
\figcaption[f1.eps]
{AB Aur NIRSPEC/Keck S(9) and TEXES/Gemini S(4), S(2), S(1) data from November 2006 observations 
overplotted with 2-component model fit. 
The dashed line shows the stellar velocity \citep{thi01}.
The increased noise in the \sone ~spectrum blueward of the position of the \sone ~line is caused by a telluric feature.
\label{gem-abaur-s9s4s2s1}
}
\end{figure}

\begin{figure}
\plotone{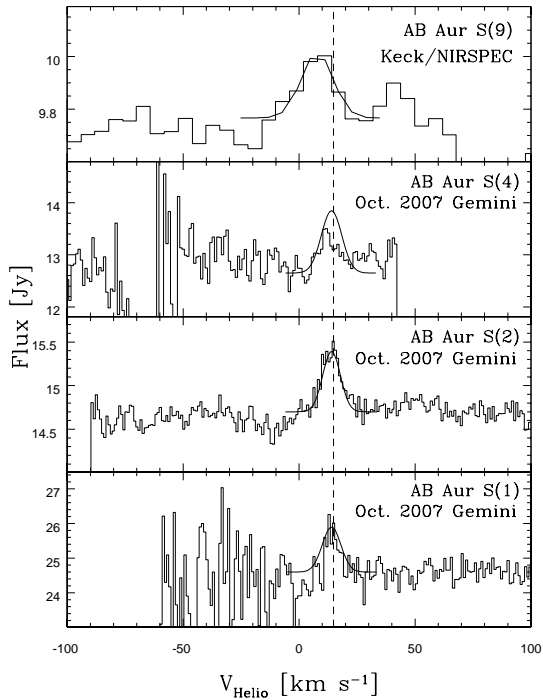}
%\centerline{f2.eps}
\figcaption[f2.eps]
{AB Aur NIRSPEC/Keck S(9) and TEXES/Gemini S(4), S(2), S(1) data from October 2007 observations 
overplotted with the 2-component model fit derived using November 2006 data.
The S(1) and S(2) lines are consistent with our November 2006 observations, however the S(4) 
line appears weaker. 
The dashed line shows the stellar velocity \citep{thi01}.
The increased noise in the \sone ~spectrum blueward of the position of the \sone ~line is caused by a telluric feature.
\label{gem-abaur-oct07-s9s4s2s1}
}
\end{figure}

\subsection{Individual Sources with H$_{2}$ Detections}
\subsubsection{AB Aur}
AB Aur is a Herbig Ae star surrounded by circumstellar material extending 
to at least $r \sim 450$ AU \citep{mannings97}.  
It has a spectral type A0 \citep{hernandez04} and is located 144 pc 
away \citep{vanden98}.
The disk surrounding the 2.4 M$_\odot$ central star \citep{vanden98} has 
an estimated mass of 0.013 M$_\odot$ and an inclination of $17 ^{+6}_{-3}$ deg 
as determined by modeling molecular line emission at millimeter wavelengths 
\citep{semenov05}. 
Observations by \citet{chen03} show that AB Aur is variable 
in the mid-infrared.

\citet{bitner07} described observations of the \sone, \stwo, and \sfour ~H$_{2}$ 
lines using TEXES.  
A single-temperature LTE model fit to the lines yielded a temperature of 
T $= 670$ K and M $=0.52$~M$_{\oplus}$ for the emitting gas.  
In this paper, with the addition of observations 
of the \snine ~line, gas at a single 
temperature no longer fits the data.  
The two-temperature fit shows that essentially all the \snine ~flux comes 
from a very small amount of gas ($\approx$ 0.075 $M_\oplus$) at a temperature
likely close to the dust sublimation temperature 
(Figure~\ref{gem-abaur-s9s4s2s1}).
It also appears that the high temperature component is slightly blue shifted
relative to the gas responsible for the lower energy lines. 
We observed AB Aur from Gemini on two separate occasions, once during 2006 November 
and again in 2007 October (Figures~\ref{gem-abaur-s9s4s2s1} 
and~\ref{gem-abaur-oct07-s9s4s2s1}).
In Figure~\ref{gem-abaur-oct07-s9s4s2s1} we show the observations from 2007 
overplotted with the two-temperature model based on the 2006 data.
While the model is a reasonable fit to the \sone ~and \stwo ~lines, 
the \sfour ~line appeared weaker in our 2007 observations.
Table~\ref{tab:lines} shows details of individual Gaussian fits to each of 
the lines from both years.  
The \stwo ~line fluxes are in excellent agreement while the \sone ~and 
\sfour ~observations show differences.
It is possible that the source varied between the two observations.
However, the agreement among the two different \stwo ~observations combined 
with the fact that we are more sensitive there than at \sone ~or \sfour ~casts 
doubt on this possibility. 

Modeling of the FWHM of the \stwo ~line profile accounting for Keplerian, instrumental, 
and thermal broadening suggests that the emission arises near 18 AU in the disk 
\citep{bitner07}.
Spatially resolved mid-infrared continuum images at 11.6 $\mu$m using Michelle 
on Gemini North by \citet{marinas06} reveal a source size of 17 AU for the 
emission that remains after the subtraction of a comparison PSF star.
Comparison with the passive flared disk model of \citet{dullemond01} suggests 
that, as the disk emerges from the shadow of the inner rim near 10 AU, 
dust grains in the surface layer are heated by direct stellar radiation, which 
produces the mid-IR continuum flux.
Their derived average dust temperature in this region of $\sim 200$ K is significantly 
lower than the gas temperature based on the H$_{2}$ observations, implying an additional 
heating source for the gas.
Likely candidates for the source of additional heating are X-ray, UV, and accretion heating.
\citet{roberge01} detected H$_{2}$ in absorption towards AB Aur, probably located 
in the envelope around the star.
They derive a temperature and column density of T = 212 K and 
N(H$_{2}$) = $6.8\times10^{19}$ cm$^{-2}$ for the absorbing gas.
If we assume the emission is spread evenly over our TEXES beam, the flux in our lines 
due to this gas is less than 10$^{-16}$ \escm, significantly lower than our detected fluxes.
\citet{brittain03} detected CO fundamental rovibrational emission from AB Aur and 
found that the emission was coming from both hot (1540 K) CO in the inner disk rim and 
cold (70 K) gas farther out in the flared region of the disk.
Our derived gas temperature based on the H$_{2}$ lines falls between the hot and cold 
components seen in the CO observations, suggesting we are not seeing the same gas. 

\begin{figure}
\plotone{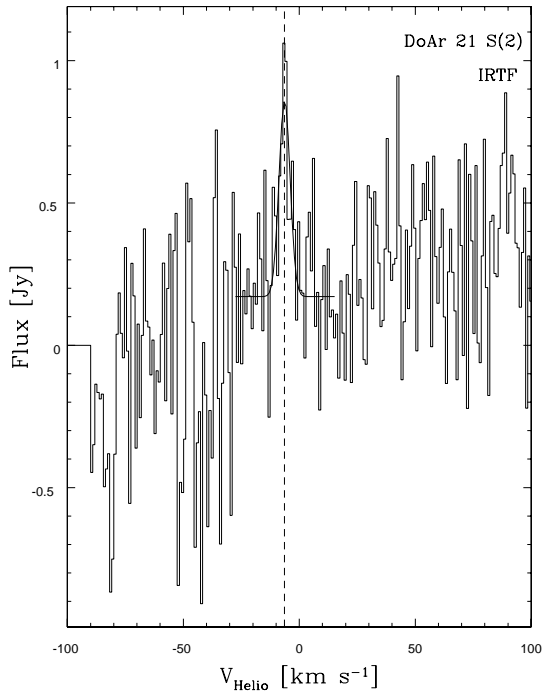}
%\centerline{f3.eps}
\figcaption[f3.eps]
{DoAr 21 TEXES/IRTF S(2) overplotted with Gaussian fit.  The dashed line indicates the systemic velocity of the 
$\rho$ Oph cloud region \citep{loren90}. 
\label{irtf-do21-s2}
}
\end{figure}

\subsubsection{DoAr 21}
DoAr 21 is located in the $\rho$ Ophiuchus molecular cloud and is classified 
as a weak-line T Tauri star \citep{bouvier92}.
The distance to $\rho$ Ophiuchus remains a source of debate with distance estimates 
ranging between 119 and 165 pc \citep{mamajek08,lombardi08,loinard08}.
For the sources in our sample, we adopt the recent distance estimate by \citet{lombardi08} 
of 119 pc.
Continuum observations reveal only a slight infrared excess \citep{wilking01} 
and millimeter observations suggest the amount of circumstellar dust is less than 
10$^{-6}$ M$_\odot$ corresponding to a gas mass less than 10$^{-4}$ M$_\odot$ assuming 
the standard gas-to-dust ratio \citep{andre94}. 
Any circumstellar disk around the star is tenuous.
\citet{bary02} detected narrow, 9 km s$^{-1}$ FWHM, near-infrared H$_{2}$ 
\textit{v}=1-0 \sone ~emission centered at the systemic velocity of DoAr21 and 
concluded that the line arose from a circumstellar disk
within $\sim$110 AU of the star.
Due to the absence of a double-peaked line profile, \citet{bary02} concluded that 
the circumstellar disk in the system has an inclination $<$45$^{\circ}$.
\citet{bary03} calculated that the line emission could be produced by a mass of
$2.7 \times 10^{-2}$ M$_{\oplus}$ H$_{2}$ gas in LTE at T = 1500 K 
located in a thin surface layer of the disk.
This mass was calculated assuming a distance of 160~pc.  Correcting
their mass to the 119~pc we assume for DoAr 21, this gas 
would produce a \stwo ~line flux at 12 $\mu$m of $0.6 \times 10^{-15}$ 
ergs s$^{-1}$ cm$^{-2}$, smaller than both our detection from the IRTF 
and the upper limit based on Gemini data.

We observed DoAr 21 with TEXES on both the IRTF and Gemini North telescopes.
From the IRTF, we detected narrow, FWHM $\sim$6 km s$^{-1}$, H$_{2}$ \stwo 
~emission (Figure~\ref{irtf-do21-s2}), 
while the line was not seen in our Gemini observations.
Since our slit size on the sky is smaller on Gemini than the IRTF, a plausible 
explanation for the discrepancy is that the emission is spatially outside our 
Gemini slit.
At the H$_{2}$ \stwo ~setting, our slit widths are 1.4\arcsec ~and 0.54\arcsec ~on the IRTF and 
Gemini respectively.
At an adopted distance of 119~pc to DoAr 21, these slit widths translate to physical 
distances from the central source of 83 AU and 32 AU.
Our data suggest that the observed H$_{2}$ emission arises outside of the inner 
32 AU around DoAr 21.
The rovibrational emission seen in DoAr 21 \citep{bary02} was observed 
using the Phoenix spectrometer \citep{hinkle98} on the NOAO 4-meter telescope 
on Kitt Peak with a seeing-limited slit width, the same as TEXES
at the IRTF.
If our observations probe the same gas, this is consistent with the possibility 
that the emitting region is outside of the inner $\sim$30 AU. 

\begin{figure}
\plotone{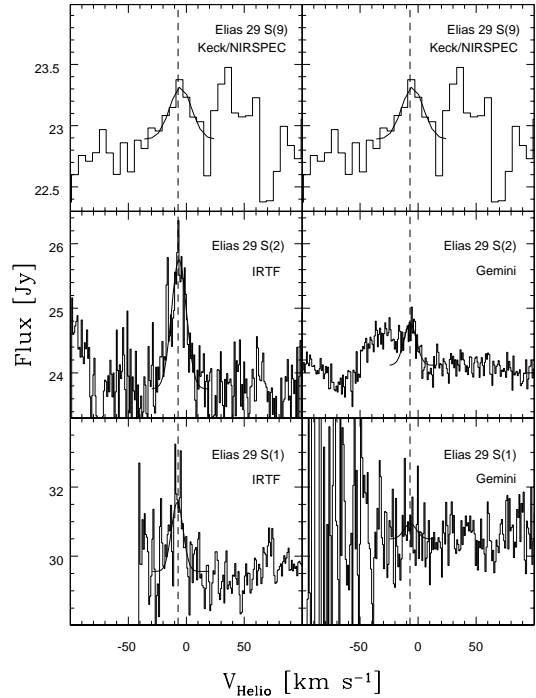}
%\centerline{f4.eps}
\figcaption[f4.eps]
{Elias 29 NIRSPEC/Keck S(9), TEXES/IRTF S(2), S(1), and TEXES/Gemini S(2), S(1) overplotted with 1-component model fit. 
The dashed line indicates the systemic velocity of Elias 29 \citep{boogert02a}.  The continuum 
has been scaled to agree with the \textit{ISO} SWS measurement.
The bump redward of the \snine ~feature is due to telluric contamination.
Poor telluric division of a water feature is apparent on the blue side of the Gemini \stwo ~line.
\label{el29-s9s2s1}
}
\end{figure}

\subsubsection{Elias 29}
Elias 29 is a class I protostar located in the $\rho$ Ophiuchus molecular cloud.
Along with several nearby protostars, it is located along a dense ridgelike 
structure seen in HCO$^{+}$ 3--2 emission \citep{boogert02a}.
The central source is surrounded by a circumstellar disk and a remnant envelope
several times more massive than the disk \citep{lommen08}.
Modeling of the spectral energy distribution \citep{boogert02a} constrains the disk 
size to $\sim$500 AU with an inclination less than 60$^{\circ}$ and mass of 
0.012 M$_\odot$.

Using the \textit{Infrared Space Observatory (ISO)} Short Wavelength Spectrometer, 
\citet{ceccarelli02} saw evidence for a disk around Elias 29 with a superheated 
surface layer and mass similar to disks around Herbig AeBe stars.
These authors suggested that Elias 29 may actually be a deeply embedded Herbig AeBe star.
\citet{boogert02a} also suggested Elias 29 might be a heavily
extincted T Tauri or Herbig AeBe star.
Elias 29 drives a bipolar CO outflow \citep{bontemps96} with velocities approaching 
$\sim 80$ km s$^{-1}$ \citep{boogert02a}.
Knotty H$_{2}$ 1-0 \sone ~emission suggests the presence of a precessing jet interacting 
with the surrounding medium and clearing the protostellar envelope \citep{ybarra06}.

We used TEXES on the IRTF in 2003 June and on Gemini North in 2006 July to observe 
Elias 29 at the H$_{2}$ \sone ~and \stwo ~settings.
We obtained observations of the H$_{2}$ \snine ~line taken with Keck/NIRSPEC during 
several runs between 2000 and 2005.
The data are shown overplotted with a single temperature LTE model fit in 
Figure~\ref{el29-s9s2s1}. 
The Gemini spectrum shows poor telluric division on the blue
side of the \stwo~line.  A telluric water feature at this velocity apparently
was not well corrected.
The results of the single temperature LTE model fits to the data are listed in 
Table~\ref{tab:ltegemini}.
The data are well fit by emission from less than 1 M$_{\oplus}$ of gas at T$\sim$1000 K.
The \stwo ~line was $>$3 times stronger in our IRTF observations compared to Gemini and the 
\sone ~line is clearly detected in our IRTF observations but not seen from Gemini.
There are two possible explanations for these discrepancies.  
Either the H$_{2}$ line flux actually changed between our observations from the IRTF 
and Gemini, or the emission is spatially extended and our wider slit on the IRTF 
took in more of the line flux.
We saw no evidence of spatially resolved H$_{2}$ emission along the slit of our Gemini 
observations at scales of 0.4 arcsec ($\sim$25 AU in radius at 119 pc).
The continuum level did not vary significantly between our IRTF and Gemini observations.

\begin{figure}
\plotone{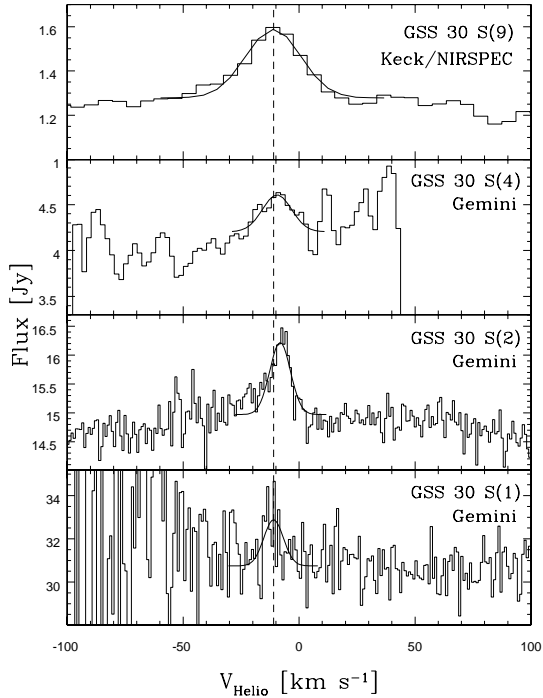}
%\centerline{f5.eps}
\figcaption[f5.eps]
{GSS 30 NIRSPEC/Keck S(9) and TEXES/Gemini S(4), S(2), S(1) overplotted with 2-component model fit. 
The dashed line shows the systemic velocity \citep{pontop02}.  The continuum has been scaled to 
agree with the \textit{Spitzer} IRS measurement.
\label{gem-gss30-s9s4s2s1}
}
\end{figure}

\subsubsection{GSS 30 IRS 1}
GSS 30 IRS 1 is a class I source \citep{wilking89} in the $\rho$ Ophiuchus 
molecular cloud.
Near-infrared polarimetry by \citet{chrys97} suggests that the source is surrounded 
by a large flattened dusty envelope and a more compact circumstellar disk.
Modeling of near-infrared polarimetry data suggests the disk has an inclination 
of $\sim$65$^{\circ}$, i.e. closer to edge-on \citep{chrys96}.
Nearby molecular outflow activity has been observed but is not clearly associated 
with GSS 30 IRS1 itself.
\citet{tamura90} observed high velocity millimeter CO emission to the south of 
GSS 30 IRS1 that is likely associated with the nearby VLA 1623 jet \citep{andre90}.

\citet{pontop02} described observations of fundamental rovibrational CO emission 
at 4.7 $\mu$m.
The lines are unresolved at R = 5000 and are spatially extended up to 320 AU 
from the central source.
The authors proposed that the line emission arises from post-shocked gas from the 
inner region of the circumstellar disk, which is then scattered into our line of 
sight by the surrounding envelope.
The lines are consistent with 1-100 M$_{\oplus}$ of gas in LTE at T $=515$ K 
with a spatial extent of $\sim$20-100 AU.

Our pure rotational H$_{2}$ data cannot be fit by a LTE single temperature/mass 
model, but is fit reasonably well with a two component model 
(Figure~\ref{gem-gss30-s9s4s2s1}).
Using a two component fit, we find a temperature for the low J lines nearly 
identical to \citet{pontop02}.
The \snine ~emission arises in a small amount of hot gas, significantly
hotter at 3300~K than the dust evaporation temperature, and has a broader 
line width than the other H$_{2}$ lines.
Because such hot gas temperatures are only effectively constrained 
by the \snine ~line, our determination of this temperature is more 
uncertain.
Since the H$_{2}$ molecule may form in a high rotational level \citep{wagenblast92}, 
this apparent high temperature, if valid, may be due to H$_{2}$ formation pumping.
As noted by \citet{pontop02}, the intrinsic line widths are expected to be 
$\sim$10 km s$^{-1}$ or larger in the case where a dissociating accretion shock is 
responsible for the emission.
The width of the low J pure rotational H$_{2}$ lines are consistent with this model.
The hot gas, however, is likely located in a separate location.

\begin{figure}
\plotone{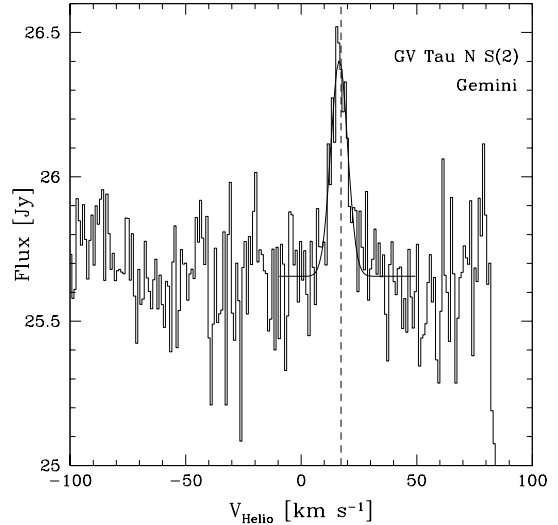}
%\centerline{f6.eps}
\figcaption[f6.eps]
{GV Tau N TEXES/Gemini S(2) overplotted with Gaussian fit.  
The dashed line indicates the systemic velocity of the GV Tau system \citep{hoger98}. 
\label{gem-gvtau-s2}
}
\end{figure}

\subsubsection{GV Tau N}
GV Tau is a pre-main-sequence binary system in the L1524 molecular cloud.
The two components are separated by 1.2\arcsec ~(170 AU at 140 pc) \citep{leinert89}.
The southern component is optically visible while GV Tau N is heavily extincted.
Near-infrared imaging and optical polarimetry show that the system is surrounded 
by a flattened, edge-on circumbinary envelope out to $\sim$1000-1500 AU \citep{menard93}.
GV Tau N shows near-infrared variability on timescales as short as a month, which 
has been attributed to clumpiness in the surrounding material \citep{leinert01}.
The GV Tau system has a spectral energy distribution rising through the mid-infrared 
leading to its classification as a class I source, however, millimeter observations by 
\citet{hoger98} suggest that most of the envelope has disappeared.
GV Tau N appears to be driving a Herbig-Haro outflow \citep{devine99}.
\citet{gibb07} reported the detection of near-infrared absorption lines due to CO, HCN, 
and C$_{2}$H$_{2}$ toward GV Tau S and derived a CO rotational temperature of 
$\sim$200 K, suggesting the observed gas is in the inner region of a circumstellar disk.
\citet{doppmann08} observed HCN, C$_{2}$H$_{2}$, and CO absorption towards GV Tau N 
but did not detect molecular absorption towards GV Tau S.
Mid-infrared observations with TEXES on Gemini in 2006 November also showed HCN 
absorption towards GV Tau N but not GV Tau S (Najita et al., in preparation).
  
We observed GV Tau N at the \stwo ~setting (Figure~\ref{gem-gvtau-s2}) 
from Gemini and detected emission at a flux level and line width consistent 
with other detections in our sample.
The detection of just a single line precludes an estimate of the temperature 
and mass of the emitting gas.
\citet{doppmann08} observed near-infrared H$_{2}$ emission from GV Tau N 
in the \textit{v}=1-0 \stwo ~and \textit{v}=1-0 \szero ~lines.
The line centroids are consistent with our mid-infared \stwo ~detection and with 
the systemic velocity of GV Tau.
The \textit{v}=1-0 \stwo ~line emission is stronger than the \textit{v}=1-0 
\szero ~by a factor that is consistent with the shock excitation seen in classical 
T Tauri stars by \citet{beck08}.

\begin{figure}
\plotone{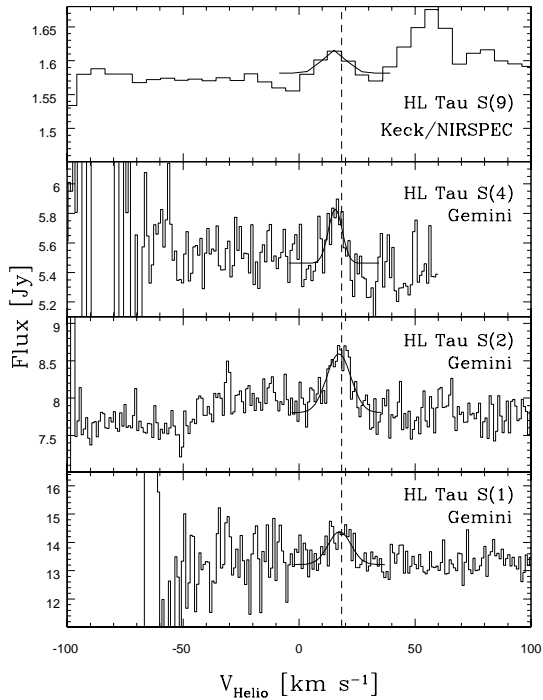}
%\centerline{f7.eps}
\figcaption[f7.eps]
{HL Tau NIRSPEC/Keck S(9) and TEXES/Gemini S(4), S(2), S(1) overplotted with 2-component fit. 
The dashed line shows the stellar velocity \citep{white04}.  The continuum has been scaled to 
agree with the \textit{ISO} SWS measurement.
The bump redward of the very weak \snine ~feature is due to telluric contamination.
\label{gem-hltau-s9s4s2s1}
}
\end{figure}

\subsubsection{HL Tau}
The young stellar object, HL Tau, is in an intermediate stage between an embedded 
class I protostar and an optically visible T Tauri star \citep{pyo06}.
It is surrounded by a 0.05-0.07 M$_{\odot}$ circumstellar disk with an outer radius 
of 90-160 AU \citep{mundy96} inclined by 66-71 degrees from face-on \citep{lucas04}.
Molecular carbon absorption suggests the presence of an infalling envelope 
\citep{grasdalen89}.
A collimated jet is seen from HL Tau in [Fe II] 1.6 $\mu$m emission surrounded 
by a wide-angled wind, which produces shocked near-infrared H$_{2}$ emission \citep{takami07}.
Spatially resolved observations by \citet{beck08} with NIFS on Gemini North confirm 
that the near-infrared H$_{2}$ \textit{v}=1-0 \sone ~emission is consistent with the 
location of the jet associated with the system.
Nearly half of the NIR H$_{2}$ emission seen by \citet{beck08} is spatially coincident 
with the continuum.
If the spatial distribution of the mid-infrared H$_{2}$ emission is the same 
as the near-infrared H$_{2}$ emission, our observed emission lines may also originate in 
circumstellar gas shocked by the jet. 
Broad near-infrared CO emission from the hot (T $\sim$ 1500 K) inner disk along 
with narrow CO absorption likely originating in the outer flared disk are observed 
from HL Tau \citep{brittain05}.
\citet{brittain05} estimated that the emission arises from CO with a column density 
of $4 \times 10^{16}$ cm$^{-2}$ in a region between 0.066 AU and 0.53 AU.
This translates to $\sim 10^{-4}$ M$_{\oplus}$ of H$_{2}$ at 1500 K, which would 
produce pure rotational mid-infrared H$_{2}$ emission at levels $\sim 10^{-18}$ ergs 
s$^{-1}$ cm$^{-2}$, well below our detection limits.
This combined with the average FWHM of the CO lines observed by \citet{brittain05} 
(45 km s$^{-1}$) makes clear that our observations are probing gas at larger radii.
From the IRTF, we detected emission in the H$_{2}$ \stwo ~line, while from Gemini 
we detected H$_{2}$ \sone, \stwo, and \sfour ~emission.
A single temperature LTE model comprised of $\sim$1 M$_{\oplus}$ of 465 K gas fits 
the observations well (Figure~\ref{gem-hltau-s9s4s2s1}).
The emission lines are all narrow with FWHM near 10 km s$^{-1}$.

\subsection{Putting TEXES Results into the Context of Other Circumstellar Gas Observations}
Emission has been detected in the near-infrared \textit{v}=1-0 S(1) transition of H$_{2}$ 
from several T Tauri stars \citep{bary03}.
The detected emission shares similar characteristics with our observations 
of the mid-infrared H$_{2}$ lines.
In both cases the lines are narrow and centered at the systemic velocity of the star.
The observed near-infrared H$_{2}$ lines all have FWHM $\sim$10 km s$^{-1}$,
suggesting that the emission arises 10-50 AU away from the star and is possibly 
the result of excitation by UV or X-rays.
The spatially resolved observations of \textit{v}=1-0 S(1) H$_{2}$ emission from the 
circumstellar environments of classical T Tauri stars by \citet{beck08} are most 
consistent with shock excited emission from outflows rather than UV or X-ray excitation.

Three of the sources with detected near-infrared H$_{2}$ emission in the \cite{bary03} 
sample were also observed as part of our TEXES H$_{2}$ mid-infrared survey.
Two of those sources, GG Tau and LkCa 15, do not have detectable levels of mid-infrared 
H$_{2}$ emission while the other, DoAr 21, was detected in the \stwo ~line in our IRTF 
observations but not when re-observed from Gemini.
\citet{bary03} computed the amount of H$_{2}$ gas required to produce the observed 
emission in these sources under the assumption of LTE at T=1500 K.
Based on their derived gas masses for these three sources, we computed the amount of emission 
we would see in the mid-infrared H$_{2}$ \sone ~and \stwo ~lines.
For GG Tau and LkCa 15, the predicted line fluxes are all less than a few times 
10$^{-16}$ \escm\ consistent with our derived upper limits.
In the case of DoAr 21, the predicted line fluxes are less than $10^{-15}$ 
ergs s$^{-1}$ cm$^{-2}$, smaller than both our non-detections from Gemini 
and the detected \stwo ~line flux observed with the IRTF from DoAr 21. 

CO fundamental rovibrational emission near 4.6 $\mu$m is detected in many T Tauri 
stars \citep{najita03} and Herbig AeBe stars \citep{blake04}.
The A-values for the CO fundamental lines are much larger than those of the pure rotational 
mid-infrared H$_{2}$ lines and so, assuming LTE, the CO lines are more sensitive to 
small column densities of gas.
The lines are broad and centrally peaked with FWHM of 50-100 km s$^{-1}$, 
suggesting the emission arises from $\lesssim$ 0.1 AU to 1-2 AU \citep{najita07a}.
The CO near-infrared observations typically reveal temperatures of 1000-1500 K and 
CO column densities of $\sim$10$^{18}$ cm$^{-2}$ \citep{najita07a}.
Assuming the CO fundamental emission arises from 0.1-2 AU with a CO column 
density of 10$^{18}$ cm$^{-2}$ and a CO/H$_{2}$ ratio of $2.7 \times 10^{-4}$ \citep{lacy94}
gives a mass of H$_{2}$ in this region of $\sim$10$^{-2}$ M$_{\oplus}$.
The line fluxes in the mid-infrared H$_{2}$ lines from such gas at T = 1000 K are 
less than 10$^{-15}$ ergs s$^{-1}$ cm$^{-2}$, smaller than our detection limits.
The additional line broadening due to the larger rotation speeds in this part of the 
disk would further decrease the chances of this gas being seen in 
our high resolution mid-infrared H$_{2}$ observations.
HD 141569 is a unique source among those with observed near-infrared CO emission in that 
the emission arises from gas at a much cooler temperature (190 K) and at a location 
$\gtrsim$ 17 AU in the disk \citep{brittain03}.
\citet{brittain03} derived a mass for the emitting CO gas of 10$^{19}$ g 
($\sim$10$^{-9}$ M$_{\oplus}$).
For a rotational temperature of 190 K, assuming the CO/H$_{2}$ ratio derived by \citet{lacy94}, 
the expected mid-infrared H$_{2}$ line fluxes are smaller than 10$^{-20}$ 
ergs s$^{-1}$ cm$^{-2}$, consistent with our non-detections.

\subsection{Location of Emitting H$_{2}$ Gas and Possible Excitation Mechanism}
The six stars with detected H$_{2}$ emission all have narrow, spectrally resolved 
line widths between 7 and 15 km s$^{-1}$.
Added to the fact that the line fluxes are all similar and that the lines are 
centered at the stellar velocity, this suggests that the excitation mechanism 
is similar in each case.
If the emission originates in a circumstellar disk, our spectrally resolved lines allow 
for the determination of the approximate emission radius.
We created simple models for the line widths originating from a Keplerian disk that contributes 
equally at all radii within some annulus and convolved these line profiles with the TEXES 
instrumental profile and thermal broadening appropriate for gas at T = 500 K, roughly 
the temperature derived from our observations.
For a star of one solar mass with a disk inclination angle of 45$^{\circ}$, the range of 
line widths detected in our sample corresponds to emission from disk radii between 
10 and 50 AU.

In all cases, our observations show the emission is spatially unresolved along the slit of 
our Gemini observations at scales of 0.4 arcsec and coincident with the source continuum.
This combined with the fact that the lines are spectrally resolved with FWHM $\sim$10 km s$^{-1}$ 
suggests that we are not seeing emission from an extended envelope surrounding the source, 
which would produce narrow, spectrally unresolved lines.  
Furthermore, it is unlikely we are seeing the results of shocks 
associated with jets and outflows commonly seen near young stars, as
that would produce broader lines with some displacement from the stellar
velocity.

However, the spatially resolved ($\sim$0.1\arcsec) observations of near-infrared H$_{2}$ emission from the 
circumstellar environments of six classical T Tauri stars by \citet{beck08} demonstrate 
that we cannot rule out shocked emission as the source of our observed mid-infrared lines 
in all cases.
One source we share with \citet{beck08} is HL Tau.
\citet{beck08} concluded that their spatially resolved observations 
of the near-infrared \textit{v}=1-0 \sone ~H$_{2}$ line from HL Tau are consistent 
with the location of the jet in the system.
In addition, nearly half of the observed H$_{2}$ emission from HL Tau is spatially 
coincident with the continuum and the line centroid is within 10 km s$^{-1}$ of the stellar 
velocity.
Assuming the spatial position of the near-infrared and mid-infrared H$_{2}$ emission 
is the same, this suggests that the emission we observed may also arise in shock excited 
circumstellar gas.
However, there is a notable difference in the overall results of the spatially resolved, near-infrared 
H$_{2}$ observations by \citet{beck08} and our mid-infrared H$_{2}$ observations. 
In half of the six stars observed, the near-infrared line centroids differed by more than 
10 km s$^{-1}$ from the stellar velocity whereas in our sample, all of the H$_{2}$ line 
centroids are within a few km s$^{-1}$ of the stellar velocity.
  
The results of spectral energy distribution modeling of the dust temperature in the surface 
layer of a disk around a typical T Tauri star show 
that the dust temperature is $\lesssim$ 200 K at disk radii larger than 10 AU where 
our H$_{2}$ emission arises \citep{chiang97}.
Stellar heating of dust grains in the disk atmosphere coupled 
to the gas temperature through gas/grain collisions is insufficient to explain 
the high gas temperatures derived from our observations.
\citet{carmona08a} computed the expected emission in the H$_{2}$ \sone ~and \stwo 
~lines from the optically thick, two-layer disk model of \citet{chiang97} in which 
dust in the disk surface layer absorbs stellar radiation and heats the gas.
They found that the amount of gas in the warm surface layer of the disk 
is less than a few Earth masses.
At a distance of 140 pc, this leads to predicted mid-IR H$_{2}$ line fluxes of 
$10^{-17}-10^{-16}$ ergs s$^{-1}$ cm$^{-2}$, much lower than both the levels 
of our detections and our upper limits.
\citet{carmona08a} pointed out that the two-layer approximation to the disk structure 
leaves out details that could significantly contribute to H$_{2}$ emission.
They found that departures from thermal coupling between gas and dust 
in the disk surface layer as well as larger than interstellar gas-to-dust ratios 
can lead to detectable levels of H$_{2}$ emission.
Two plausible mechanisms for additional gas heating in the surface layers of disks 
are accretion shocks due to infalling matter onto the disk and X-ray/UV irradiation.
That accretion onto the disk may play a role in exciting H$_{2}$ emission is consistent with 
the preferential detection of H$_{2}$ emission from the class I sources in our sample, 
which possess a surrounding envelope of material in addition to a disk.

\citet{neufeld94} have calculated the physical and chemical structure of 
shocks resulting from accretion onto a circumstellar disk, but they did not 
publish the strength of the pure rotational H$_{2}$ lines from their model. 
However, we have reviewed the results of unpublished models from their study 
to determine the parameters that would produce H$_{2}$ emission at the levels detected 
in our observations.
\citet{neufeld94} only considered preshock densities above $3\times10^{7}$ cm$^{-3}$.
However, below this density, the fraction of the total cooling from H$_{2}$ emission increases.
A shock with a preshock density of 10$^{6}$ cm$^{-3}$ striking the disk at 5 km s$^{-1}$ at 30 AU 
implies an accretion rate of $5\times10^{-8}$ M$_\odot$ yr$^{-1}$.
The resulting emission in the \sone, \stwo, and \sfour ~H$_{2}$ lines is $\sim$10$^{-6}$ L$_\odot$.
This accretion rate is similar to the measured rates for the sources where we have 
detected H$_{2}$ emission and these line luminosities are only slightly lower than 
our observations.
One way to increase the H$_{2}$ cooling in the model is to suppress the cooling in 
H$_{2}$O lines in the shock by assuming that H$_{2}$O is frozen out as water ice 
in the preshock gas. 
It is plausible that the H$_{2}$O would be frozen out for the 5 km s$^{-1}$ shocks 
since they are slow enough that they would not return the H$_{2}$O to the gas phase.
If this were the case, the luminosity in our H$_{2}$ lines would increase by an order 
of magnitude to about 10$^{-5}$ L$_\odot$.
The line luminosities detected in our sample range from 10$^{-6}$ to 10$^{-5}$ L$_\odot$ 
suggesting that shock heating due to accretion onto a disk is a plausible excitation 
mechanism. 
Additional details including predicted CO line strengths can be found in the Appendix.

\citet{nomura07} modeled the molecular hydrogen emission from a disk surrounding a 
typical T Tauri star, taking into account the heating of gas by X-ray and UV irradiation 
from the central star. 
The resulting gas temperature in the surface layer of the disk is much higher 
than the dust temperature.
X-ray heating dominates in the inner region and surface layer of the disk.
At 10 AU, the disk surface temperature reaches over 1000 K.
Even at 100 AU, the temperature in the disk surface layer reaches 200 K.
These temperatures combined with the fact that the gas is hotter than the dust 
creates favorable conditions for the production of mid-infrared H$_{2}$ 
emission lines.
The predicted line fluxes derived by \citet{nomura07} appropriate for TW Hya
vary depending on the adopted dust size distribution.
The quoted line fluxes at our wavelengths are $4.4-9.9 \times 10^{-15}$ ergs 
s$^{-1}$ cm$^{-2}$ at \sone, $1.7-4.8 \times 10^{-15}$ ergs s$^{-1}$ cm$^{-2}$ 
at \stwo, and  $0.4-8.8 \times 10^{-15}$ ergs s$^{-1}$ cm$^{-2}$ at \sfour.
At distances more typical of the sources in our sample, these fluxes are 
$\sim$10$^{-16}$ ergs s$^{-1}$ cm$^{-2}$.
In the sources where we detect line emission, the derived line fluxes are 
all larger, typically 10$^{-15}$-10$^{-14}$ ergs s$^{-1}$ cm$^{-2}$.
\citet{gorti08} have also modeled line emission from the upper layers of 
optically thick disks.
Their predicted H$_{2}$ emission line luminosities are 10$^{-6}$ to 10$^{-5}$ L$_\odot$ and 
are consistent with the range of detected emission in our sample.
The difference in the predictions of the two models shows that the modeling 
is very sensitive to the input parameters.
The FUV flux assumed by \citet{gorti08} in their standard model is five times higher and the 
X-ray flux is three times larger than the values 
used by \citet{nomura07} which leads to the higher predicted H$_{2}$ line fluxes.

\begin{figure}
\plotone{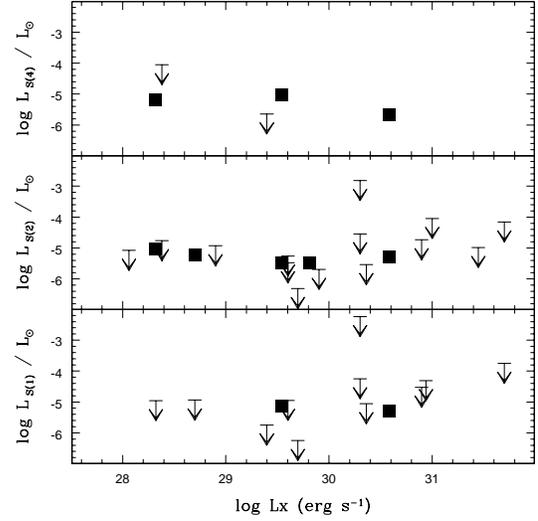}
%\centerline{f14.eps}
\figcaption[f14.eps]
{Gemini H$_{2}$ line luminosities vs. X-ray luminosity of each source.
The solid points are H$_{2}$ detections and the arrows represent upper limits.
No correlation between the presence of H${2}$ emission and X-ray luminosity is apparent.
\label{gem-xray}
}
\end{figure}

\begin{figure}
\plotone{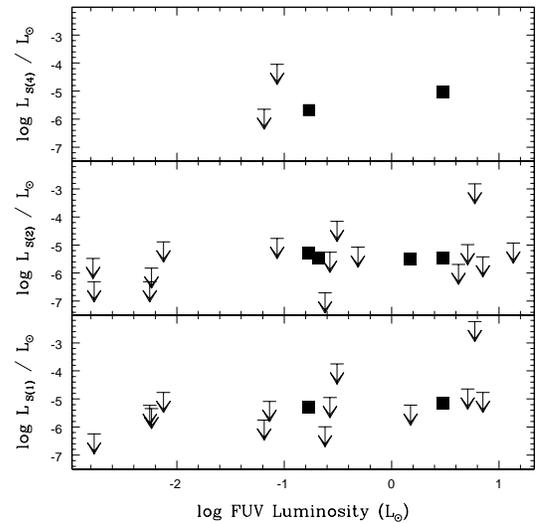}
%\centerline{f15.eps}
\figcaption[f15.eps]
{Gemini H$_{2}$ line luminosities vs. FUV luminosity of each source.
The solid points are H$_{2}$ detections and the arrows represent upper limits.
There is a hint of a correlation between FUV luminosity and detected H$_{2}$ emission 
for the small sample of \sfour ~data points, however there is no clear overall correlation.
\label{gem-fuv}
}
\end{figure}

\begin{figure}
\plotone{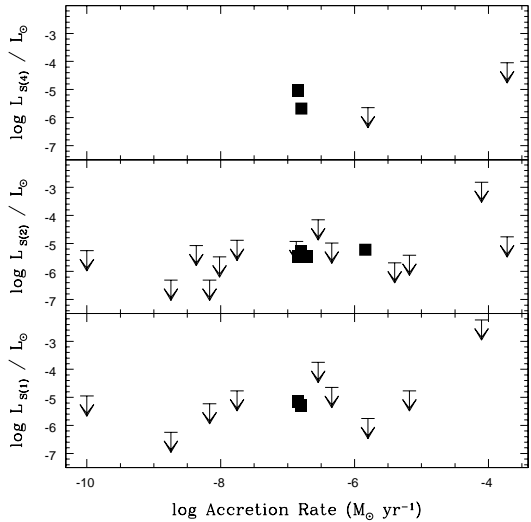}
%\centerline{f16.eps}
\figcaption[f16.eps]
{Gemini H${2}$ line luminosities vs. accretion rate measured in each source.
The solid points are H$_{2}$ detections and the arrows represent upper limits.
The sources with detected H$_{2}$ emission fall in the middle of the range of 
accretion rates.
\label{gem-accretion}
}
\end{figure}

In Figure~\ref{gem-xray}, we present a plot of our H$_{2}$ detections 
and upper limits versus the X-ray luminosity of the stars in our sample.
If high X-ray luminosities were responsible for heating the gas in disks 
above the dust temperature to produce detectable H$_{2}$ emission, we 
should see a correlation between detected H$_{2}$ emission and X-ray luminosity.
As seen in Figure~\ref{gem-xray}, such a correlation is not apparent in our data.
In several cases, we detect H$_{2}$ emission from sources at the faint end of 
the distribution of X-ray luminosities, while we did not detect H$_{2}$ emission 
from the most X-ray luminous stars in our sample.
It is worth noting that X-ray flux has been observed to vary by more than an order 
of magnitude on timescales of days \citep{kastner99}.
To definitively test for observational evidence of a correlation between X-ray flux 
and H$_{2}$ emission requires a series of coordinated observations.
UV irradiation of the disk from the central star may also contribute to heating 
the gas in the disk, leading to detectable levels of H$_{2}$ emission.
We plot H$_{2}$ line luminosity versus FUV luminosity in Figure~\ref{gem-fuv} and 
find no clear correlation, however.
Since accretion heating may play a role in producing H$_{2}$ emission, we 
plot H$_{2}$ line luminosity versus accretion rate in Figure~\ref{gem-accretion} to look 
for a correlation.
The sources in our sample with detected H$_{2}$ emission are located in the middle 
of the range of accretion rates, showing no clear correlation.
Since the dynamic range separating our H$_{2}$ detections from the non-detections 
is not very large, the lack of a correlation between L$_{x}$, L$_{FUV}$, \.M and detected H$_{2}$ 
does not conclusively rule out the possible importance of X-ray, UV, and accretion 
heating in producing H$_{2}$ emission.
We also searched for a correlation between H$_{2}$ emission and mass, age, and inclination 
but found none.

%++++++++++++++++++++++++++++++++++++++++++++++++++
%  CONCLUSIONS
%++++++++++++++++++++++++++++++++++++++++++++++++++

\section{CONCLUSIONS}

We have carried out a survey for pure rotational H$_{2}$ emission from the circumstellar environments 
surrounding a sample of 29 stars with disks and detected emission from 6.
In the case of non-detections, our upper limits constrain the amount of T $>$ 500 K gas in 
the surface layers of the circumstellar disks to be less than a few Earth masses.
Several objects in our survey have transition object SEDs implying the presence of 
an optically thin, dust-depleted inner disk: DoAr 21, GM Aur, HD 141569, and LkCa 15.
Among these sources, only DoAr 21 shows H$_{2}$ emission and it appears to be 
far from the star.
One possible explanation for transitional SEDs is grain growth \citep{strom89,dullemond05} 
whereby the inner disk becomes optically thin yet remains gas rich.
This gas could be heated through accretion as well as X-ray and UV heating.
GM Aur is of particular interest since it has an accretion rate ($\sim$10$^{-8}$ M$_\odot$ yr$^{-1}$) 
similar to typical CTTS accretion rates so should have as much gas in the inner disk as 
a typical CTTS.
If grain growth is responsible for the transitional SEDs of these sources, the available 
heating mechanisms are insufficient to produce detectable H$_{2}$ line emission.
Alternatively, grain growth may not be a good explanation for a transition 
object SED, as suggested by other demographic data \citep{najita07b}.

In all cases, the detected emission lines are narrow and centered at the stellar velocity.
The narrow range of line widths, FWHM between 7 and 15 km s$^{-1}$, along with the fact 
that the line fluxes are all similar, suggests that the mechanism for exciting the emission 
may be the same in each case.
Four of the six targets with detected emission are class I sources that show 
evidence for surrounding material in an envelope in addition to a circumstellar disk.
It is possible, and likely in the case of HL Tau, that the H$_{2}$ emission we observe 
is a result of gas in the circumstellar envelope being shock heated by an outflow.
However, the fact that all of the H$_{2}$ line centroids in our sample are within a few 
km s$^{-1}$ of the stellar velocity argues against this being the case for all of 
our detections.

Under the assumption of emission from a disk in Keplerian rotation, 
the narrow line widths imply that the emission arises at disk radii from 10-50 AU.
At such large disk radii, additional heating of the gas besides heating due to collisions with 
dust grains is required to explain the temperatures derived from our H$_{2}$ observations.
Both X-ray/UV irradiation of the disk surface layer and accretion shocks resulting from matter 
infall onto the disk are plausible candidates.
With the exception of DoAr 21, all of the sources where we detect H$_{2}$ emission 
possess both a circumstellar disk and a surrounding envelope of material.
This lends support to the possibility that the H$_{2}$ emission we observed 
may be the result of shocks in the disk due to infalling material.

Models of molecular hydrogen emission from disks that assume sufficient levels of stellar X-ray and 
UV irradiation \citep{gorti08} predict line fluxes that are consistent with 
our observations.
In contrast, models which assume smaller values of stellar UV and X-ray irradiation \citep{nomura07} 
produce weaker H$_{2}$ emission than observed in our sample.
We looked for evidence of a correlation between X-ray/UV luminosity and the presence of H$_{2}$ 
emission but found none.  
We note that the X-ray and UV luminosities used for the purpose of searching 
for a correlation with H$_{2}$ emission were not measured at the same time.
To definitively test for a correlation between X-ray/UV luminosity and the presence 
of H$_{2}$ emission will require a series of coordinated observations.

\acknowledgements
We thank the Gemini staff, and John White in particular, for their support 
of TEXES observations on Gemini North.
We are also grateful to the anonymous referee for helpful comments on this 
manuscript.
The development of TEXES was supported by grants from the NSF and the 
NASA/USRA SOFIA project.
Modification of TEXES for use on Gemini was supported by Gemini 
Observatory.  
Observations with TEXES were supported by NSF grant AST-0607312.
MJR acknowledges support from NSF grant AST-0708074 and NASA grant NNG04GG92G.
Some of the data presented herein were obtained at the W.M. Keck Observatory, 
which is operated as a scientific partnership among the California Institute 
of Technology, the University of California and the National Aeronautics 
and Space Administration.  The Observatory was made possible by the generous 
financial support of the W.M. Keck Foundation.
This work is based on observations obtained at the Gemini Observatory, 
which is operated by the Association of Universities for Research in 
Astronomy, Inc., under a cooperative agreement with the NSF on behalf 
of the Gemini partnership: the National Science Foundation (United States), 
the Particle Physics and Astronomy Research Council (United Kingdom), 
the National Research Council (Canada), CONICYT (Chile), the Australian 
Research Council (Australia), CNPq (Brazil) and CONICET (Argentina).

%++++++++++++++++++++++++++++++++++++++++++++++++++++
%  REFERENCES
%++++++++++++++++++++++++++++++++++++++++++++++++++++

\clearpage

\clearpage
\begin{deluxetable}{lccccccccc}
%preamble commands
\tabletypesize{\footnotesize}
%\rotate
\tablewidth{0pt}
%\tablenum{text}
\tablecolumns{10}
%\tableheadfrac{num}
\tablecaption{Physical properties of the stars in our sample\label{tab:props}}
\tablehead{
&
&
&
&
&
&
&
&
&
\\
\colhead{Star} &
\colhead{RA} &
\colhead{Dec} &
\colhead{Sp.T.} &
\colhead{Class} &
\colhead{d} &
\colhead{Age} &
\colhead{log L$_{x}$} &
\colhead{L$_{FUV}$\tablenotemark{a}} &
\colhead{\.M}
\\
&
&
&
&
&
\colhead{(pc)} &
\colhead{(Myr)} &
\colhead{(erg s$^{-1}$)} &
\colhead{(L$_{\odot}$)} &
\colhead{($10^{-7}~ M_{\odot}~ yr^{-1}$)}
}

\startdata
%table data
49 Ceti         &1$^{h}$34$^{m}$37$^{s}$.9 &-15$^{\circ}$40$^{\prime}$35$^{\prime\prime}$.5         &A1         &debris&61$^{1}$&7.8$^{2}$&... &0.24$^{3}$&...\\
51 Oph          &17$^{h}$31$^{m}$25$^{s}$.0 &-23$^{\circ}$57$^{\prime}$45$^{\prime\prime}$.5        &A0         &HAeBe&131$^{4}$&200$^{5}$,0.3$^{6}$&28.9$^{7}$&13.6$^{3,8}$&1.35$^{8}$ \\    
AB Aur          &04$^{h}$55$^{m}$45$^{s}$.8 &30$^{\circ}$33$^{\prime}$04$^{\prime\prime}$.3         &A0         &HAeBe&144$^{9}$&4.6$^{2}$&29.5$^{10}$&3.04$^{3}$&1.41$^{8}$\\
AS 209          &16$^{h}$49$^{m}$15$^{s}$.3 &-14$^{\circ}$22$^{\prime}$09$^{\prime\prime}$.3        &K5         &II&119$^{11}$&...&30.4$^{12}$&0.0058$^{3,13}$&... \\    
AS 353a         &17$^{h}$56$^{m}$21$^{s}$.2 &-21$^{\circ}$57$^{\prime}$23$^{\prime\prime}$.0        &M1.5$^{14}$&II&150$^{15}$&0.2$^{16}$ &$<29.9$$^{17}$&...&39.8$^{16}$ \\ 
DoAr 21         &16$^{h}$26$^{m}$03$^{s}$.0 &-24$^{\circ}$23$^{\prime}$36$^{\prime\prime}$.9        &K0         &III&119$^{11}$&0.3$^{18}$ &31.4$^{19}$&...&... \\
Elias 29        &16$^{h}$27$^{m}$09$^{s}$.5 &-24$^{\circ}$37$^{\prime}$18$^{\prime\prime}$.8        &...        &I&119$^{11}$&0.3$^{18}$ &$<28.7$$^{17}$&1.5&14.45$^{20}$ \\   
FU Ori          &05$^{h}$45$^{m}$22$^{s}$.3 &09$^{\circ}$04$^{\prime}$12$^{\prime\prime}$.0         &G3         &fuori&500$^{21}$&...&28.4$^{22}$&0.086$^{3,23}$&1900$^{21}$ \\    
GG Tau          &04$^{h}$32$^{m}$30$^{s}$.3 &17$^{\circ}$31$^{\prime}$40$^{\prime\prime}$.7         &K6         &II&140$^{24}$&1.7$^{2}$ &$<22.0$$^{25}$&0.0075$^{3,26}$&0.175$^{26}$\\
GM Aur          &04$^{h}$55$^{m}$10$^{s}$.9 &30$^{\circ}$21$^{\prime}$59$^{\prime\prime}$.5         &K5         &II&140$^{24}$&1.8$^{2}$ &29.6$^{27}$&0.0017$^{3,26}$&0.096$^{26}$ \\
GSS 30          &16$^{h}$26$^{m}$21$^{s}$.5 &-24$^{\circ}$23$^{\prime}$07$^{\prime\prime}$.8        & ...        &I&119$^{11}$&0.3$^{18}$ &$<28.3$$^{28}$&...&... \\          
GV Tau N        &04$^{h}$29$^{m}$23$^{s}$.7 &24$^{\circ}$32$^{\prime}$57$^{\prime\prime}$.6         &K3         &I&140$^{24}$&...&29.8$^{10}$&0.21&1.95$^{29}$ \\
GW Ori          &05$^{h}$29$^{m}$08$^{s}$.4 &11$^{\circ}$52$^{\prime}$12$^{\prime\prime}$.7         &K3         &II&450$^{30}$&1.0$^{31}$ &31.7$^{27}$&0.31$^{3,31}$&2.85$^{31}$ \\    
HD 141569       &15$^{h}$49$^{m}$57$^{s}$.7 &-03$^{\circ}$55$^{\prime}$17$^{\prime\prime}$.0        &B9.5       &HAeBe&99$^{4}$  &$>10.0^{9}$ &$<28.1$$^{32}$&0.486$^{3}$&0.043$^{8}$ \\          
HD 163296       &17$^{h}$56$^{m}$21$^{s}$.3 &-21$^{\circ}$57$^{\prime}$23$^{\prime\prime}$.0        &A1         &HAeBe&122$^{4}$&6.0$^{2}$&29.6$^{32}$&0.267$^{3}$&0.001$^{33}$ \\
HL Tau          &04$^{h}$31$^{m}$38$^{s}$.5 &18$^{\circ}$13$^{\prime}$58$^{\prime\prime}$.0         &K9         &I&140$^{24}$&0.77$^{34}$&30.6$^{10}$&0.17&1.6$^{29}$ \\ 
IRAS 04278+2253 &04$^{h}$30$^{m}$50$^{s}$.7 &23$^{\circ}$00$^{\prime}$11$^{\prime\prime}$.1        &F1         &I&140$^{24}$&...&...&7.1&66.1$^{29}$ \\ 
L1551 IRS 5     &04$^{h}$31$^{m}$34$^{s}$.2 &18$^{\circ}$08$^{\prime}$05$^{\prime\prime}$.3        &G-K$^{35}$ &I&140$^{24}$&...&28.3$^{10}$&...&$<140^{36}$ \\ 
LkCa 15         &04$^{h}$39$^{m}$17$^{s}$.8 &22$^{\circ}$21$^{\prime}$03$^{\prime\prime}$.5        &K5         &III&140$^{24}$&11.7$^{2}$&$<22.6$$^{25}$&0.0056$^{37,38}$&0.068$^{39}$ \\            
Lk H$\alpha$ 225&20$^{h}$20$^{m}$30$^{s}$.8 &41$^{\circ}$21$^{\prime}$24$^{\prime\prime}$.8        &...        &HAeBe&1000$^{40}$&...&...&...&... \\ 
MWC 758         &05$^{h}$30$^{m}$27$^{s}$.4 &25$^{\circ}$19$^{\prime}$56$^{\prime\prime}$.8        &A3         &HAeBe&200$^{9}$&...&...&0.073$^{3}$&... \\
RW Aur          &05$^{h}$07$^{m}$49$^{s}$.6 &30$^{\circ}$24$^{\prime}$05$^{\prime\prime}$.4        &G5         &I&140$^{24}$&2.57$^{34}$&$<29.4$$^{41}$&0.065$^{3,13}$&15.8$^{16}$ \\                 
SVS 13          &03$^{h}$29$^{m}$03$^{s}$.6 &31$^{\circ}$16$^{\prime}$01$^{\prime\prime}$.2        &...        &I&300$^{42}$&...&$<30.3$$^{43}$&0.67&6.3$^{44}$ \\  
TW Hya          &11$^{h}$01$^{m}$51$^{s}$.9 &-34$^{\circ}$42$^{\prime}$18$^{\prime\prime}$.3       &K7         &II &51$^{45}$ &10$^{46}$ &29.7$^{47}$ &0.0017 &0.018$^{48,49}$ \\              
V892 Tau        &04$^{h}$18$^{m}$40$^{s}$.7 &28$^{\circ}$19$^{\prime}$16$^{\prime\prime}$.2        &A6         &HAeBe&140$^{24}$&...&30.9$^{10}$&0.36&... \\ 
V1057 Cyg       &20$^{h}$58$^{m}$53$^{s}$.1 &44$^{\circ}$15$^{\prime}$28$^{\prime\prime}$.6        &...        &fuori&600$^{21}$&...&$<31.0$$^{17}$&0.21$^{3,23}$&... \\
V1331 Cyg       &21$^{h}$01$^{m}$09$^{s}$.1 &50$^{\circ}$21$^{\prime}$45$^{\prime\prime}$.2        &G5         &fuori&550$^{50}$&0.8$^{23}$ &$<30.9$$^{12}$&...&... \\
VV Ser          &18$^{h}$28$^{m}$47$^{s}$.8 &00$^{\circ}$08$^{\prime}$39$^{\prime\prime}$.7        &A2         &HAeBe&245$^{51}$&...&...&5.12$^{3,8}$&4.57$^{8}$ \\         
Z CMa           &07$^{h}$03$^{m}$42$^{s}$.0 &-11$^{\circ}$33$^{\prime}$02$^{\prime\prime}$.8        &F6$^{9}$   &fuori&1150$^{52}$&...&30.3$^{32}$&5.96$^{3,53}$&790$^{21}$ \\    
         
\enddata

\tablenotetext{a}{If not otherwise noted, value derived using TW Hya IUE Spectrum then scaling by relative accretion rates.}

\tablerefs{$^{1}$\citet{jayawardhana01},$^{2}$\citet{thi01},$^{3}$\citet{valenti03},$^{4}${Hipparcos},$^{5}$\citet{vanden01},$^{6}$\citet{herbertz91},$^{7}$\citet{berghofer96},$^{8}$\citet{garcialopez06},$^{9}$\citet{vanden98},$^{10}$\citet{gudel07},$^{11}$\citet{lombardi08},$^{12}$\citet{walter81},$^{13}$\citet{valenti93},$^{14}$\citet{tokunaga04},$^{15}$\citet{prato03},$^{16}$\citet{hartigan95},$^{17}$\citet{carkner98},$^{18}$\citet{luhman99},$^{19}$\citet{imanishi02},$^{20}$\citet{natta06},$^{21}$\citet{hartmann96},$^{22}$\citet{skinner06},$^{23}$\citet{herbig06},$^{24}$\citet{elias78},$^{25}$\citet{neuhauser95},$^{26}$\citet{gullbring98},$^{27}$\citet{feigelson81},$^{28}$\citet{gagne04},$^{29}$\citet{white04},$^{30}$\citet{dolan01},$^{31}$\citet{calvet04},$^{32}$\citet{stelzer06b},$^{33}$\citet{swartz05},$^{34}$\citet{siess99},$^{35}$\citet{mundt85},$^{36}$\citet{fuller95},$^{37}$\citet{bergin04},$^{38}$\citet{espaillat07},$^{39}$\citet{hartmann98},$^{40}$\citet{marvel05},$^{41}$\citet{gahm80},$^{42}$\citet{cernis90},$^{43}$\citet{getman02},$^{44}$\citet{edwards03},$^{45}$\citet{mamajek05}, $^{46}$\citet{uchida04},$^{47}$\citet{stelzer04},$^{48}$\citet{alencar02},$^{49}$\citet{herczeg04},$^{50}$\citet{shevchenko91},$^{51}$\citet{chavarria88},$^{52}$\citet{herbst78},$^{53}$\citet{stelzer06a} } 

\end{deluxetable}

% IRTF table for detections
\clearpage
\begin{deluxetable}{lcrcccccc}
%preamble commands
\tabletypesize{\footnotesize}
%\rotate
\tablewidth{0pt}
%\tablenum{text}
\tablecolumns{9}
%\tableheadfrac{num}
\tablecaption{Summary of Line Detections \label{tab:lines}}
\tablehead{
&
&
&
&
&
&
\colhead{Line} &
\colhead{Equivalent} &
\\
\colhead{Star} &
\colhead{Instrument} &
\colhead{$\lambda$} &
\colhead{Date} &
%\colhead{F$_{\nu}$} &
\colhead{Cont.} &
\colhead{Line Flux\tablenotemark{a}} &
\colhead{Luminosity} &
\colhead{Width} &
\colhead{FWHM} 
\\
&
&
\colhead{($\mu$m)}&
&
\colhead{(Jy)} &
&
\colhead{(10$^{-6}$ L$_\odot$)} &
\colhead{(km ~s$^{-1}$)} &
\colhead{(km ~s$^{-1}$)} 
}

\startdata
%table data

AB Aur & TEXES/IRTF           & 12.279&dec02,dec03& 12.7$^{b}$ & 0.93 (0.25) & 6.0& 0.86 (0.23) & 7.0 \\
AB Aur & TEXES/Gemini         & 8.025 &nov06& 12.7$^{b}$ & 1.47 (0.34) & 9.5&0.93 (0.21) & 10.4 \\
      &                       & 12.279&nov06& 14.7$^{b}$ & 0.53 (0.07) & 3.4& 0.44 (0.06) & 8.5 \\
      &                       & 17.035&nov06& 24.6$^{b}$ & 1.10 (0.3) & 7.1& 0.76 (0.21) & 9.0 \\
AB Aur &TEXES/Gemini          & 8.025 &oct07& 12.7$^{b}$ & $<1.24^{e}$ & $<8.0$& ... & ... \\
      &                       & 12.279&oct07& 14.7$^{b}$ & 0.56 (0.07) & 3.6& 0.47 (0.06) & 9.4 \\
      &                       & 17.035&oct07& 24.6$^{b}$ & 0.57 (0.16) & 3.7& 0.29 (0.08) & 6.5 \\
AB Aur &NIRSPEC/Keck          & 4.695 &jan01-dec02& 9.6$^{b}$ & 0.90 (0.07) & 5.8& 1.02 (0.08) & 15.9 \\
DoAr 21 & TEXES/IRTF          & 12.279&jun03& 0.14$^{b}$ & 0.33 (0.09) & 1.5& 24.06 (6.46) & 5.6 \\
Elias 29 & TEXES/IRTF         & 12.279&jun03& 24.0$^{c}$ & 2.45 (0.33) & 10.9&1.27 (0.17) & 15.1 \\
         &                    & 17.035&jun03& 30.5$^{c}$ & 1.64 (0.30) & 7.3&0.94 (0.18) & 11.9 \\
Elias 29& TEXES/Gemini        & 12.279&jul06& 24.0$^{c}$ & 0.70 (0.12) & 3.1& 0.36 (0.06) & 12.7 \\
Elias 29& NIRSPEC/Keck        & 4.695 &jul00-apr05& 22.9$^{b}$ & 2.12 (0.33) & 9.4& 0.44 (0.07) & 21.8 \\
GSS 30  & TEXES/IRTF          & 12.279&jun03& 14.7$^{d}$ & 1.19 (0.25) & 5.3&0.98 (0.21) & 6.8 \\
GSS 30  &TEXES/Gemini         & 8.025 &jul06& 4.21$^{d}$ & 0.78 (0.07) & 3.5&1.50 (0.14) & 14.7 \\
        &                     & 12.279&jul06& 14.7$^{d}$ & 1.13 (0.14) & 5.0&0.95 (0.12) & 10.5 \\
GSS 30  &NIRSPEC/Keck         & 4.695 &apr02& 1.2$^{b}$ & 1.97 (0.17) & 8.7&7.38 (0.62) & 28.0 \\
GV Tau N&TEXES/Gemini         & 12.279&nov06& 25.6$^{b}$ & 0.55 (0.07) & 3.4&0.27 (0.04) & 8.5 \\
HL Tau  & TEXES/IRTF          & 12.279&dec02& 7.6$^{c}$ & 1.16 (0.18) & 7.1&1.82 (0.29) & 10.9 \\
HL Tau  &TEXES/Gemini         & 8.025 &nov06& 5.5$^{c}$  & 0.34 (0.11) & 2.1&0.50 (0.16) & 7.1 \\
        &                     & 12.279&nov06& 7.6$^{c}$  & 0.84 (0.13) & 5.1&1.36 (0.22) & 12.3 \\
        &                     & 17.035&nov06& 13.4$^{c}$ & 0.84 (0.23) & 5.1&1.07 (0.29) & 11.6 \\
HL Tau  &NIRSPEC/Keck         & 4.695 &oct01-nov03& 1.6$^{b}$  & 0.09 (0.03) & 0.6&0.27 (0.06) & 11.9 \\
\enddata

\tablenotetext{a}{In units of $10^{-14}$ ~ergs ~s$^{-1}$ ~cm$^{-2}$, Value in parentheses is 1-$\sigma$ error}   
\tablenotetext{b}{Measured value}
\tablenotetext{c}{\textit{ISO} SWS archive}
\tablenotetext{d}{\textit{Spitzer} IRS}
%\tablerefs{$^{b}$Measured value, $^{c}$ISO SWS Archive, $^{d}$Spitzer IRS }
\tablenotetext{e}{3$\sigma$ limit assuming FWHM = 10.4 km s$^{-1}$}   
\end{deluxetable}

\clearpage
\begin{deluxetable}{lccccccc}
%preamble commands
\tabletypesize{\footnotesize}
%\rotate
\tablewidth{0pt}
%\tablenum{text}
\tablecolumns{8}
%\tableheadfrac{num}
\tablecaption{Results of LTE Model Fits \label{tab:ltegemini}}
\tablehead{
&
&
&
&
&
&
&
\\
\colhead{Star} &
\colhead{Telescope} &
\multicolumn{2}{c}{One Component} &
\multicolumn{4}{c}{Two Component} 
\\
%\cline{2-3}
\cline{5-8}
&
&
\colhead{Temperature (K)} &
\colhead{Mass (M$_{\oplus}$)} &
\colhead{T$_{cold}$ (K)} &
\colhead{M$_{cold}$ (M$_{\oplus}$)} &
\colhead{T$_{hot}$ (K)} &
\colhead{M$_{hot}$ (M$_{\oplus}$)} 
}

\startdata
%table data
AB Aur\tablenotemark{a} & Gemini & 670 (40) & 0.52 (0.15) & 320 (60) & 1.65 (0.52) & 1470 (100) & 0.076 (0.01) \\
%AB Aur\tablenotemark{b} & Gemini & & & 330 (50) & 1.55 (0.43) & 1500 & 0.07 (0.01) \\
GSS 30\tablenotemark{a} & Gemini & 535 (45) & 0.78 (0.16) & 520 (60) & 0.82 (0.27) & 3330 (130) & 0.002$^{+0.002}_{-0.0007}$ \\
%GSS 30\tablenotemark{b} & Gemini & & & 420 (50) & 1.2 (0.36) & 1500 & 0.041 (0.01) \\
HL Tau\tablenotemark{a} & Gemini & 465 (20) & 1.09 (0.14) & 460 (20) & 1.11 (0.17) & 1790$^{+60}_{-440}$  & 0.001$^{+0.004}_{-0.001}$ \\
%HL Tau\tablenotemark{b} & Gemini & & & 460 (20) & 1.1 (0.15) & 1500 & 0.0025 (0.0005) \\
Elias 29\tablenotemark{b} & Gemini & 1210 (90) & 0.19 (0.03) &...&...&...&... \\
Elias 29\tablenotemark{b} & IRTF & 1000 (90) & 0.78 (0.07) &...&...&...&... \\
\enddata

\tablenotetext{a}{One component fit to S(1), S(2), and S(4). Two component fit includes S(9).}   
%\tablenotetext{b}{Two component fit to S(1),S(2),S(4),and S(9) with T$_{hot}$ fixed at 1500 K.}
\tablenotetext{b}{One component fit to S(1), S(2), and S(9).}   

\end{deluxetable}

\clearpage
\begin{deluxetable}{lrlrllll}
%preamble commands
\tabletypesize{\footnotesize}
%\rotate
\tablewidth{0pt}
%\tablenum{text}
\tablecolumns{8}
%\tableheadfrac{num}
\tablecaption{Summary of Upper Limits - IRTF/TEXES\label{tab:ulirtf}}
\tablehead{
&
&
&
&
&
&
&
\\
\colhead{Star} &
\colhead{$\lambda$} &
\colhead{Date} &
\colhead{F$_{\nu}$\tablenotemark{a}} &
&
\multicolumn{3}{c}{Mass in M$_{Jup}$} 
\\
\cline{6-8}
&
\colhead{($\mu$m)}&
&
\colhead{(Jy)} &
\colhead{Line Flux\tablenotemark{b}} &
\colhead{200 K} &
\colhead{500 K} &
\colhead{800 K} 
}

\startdata
%table data
51 Oph    &12.279  &jun03& 10.9 & $<2.2$ &$<1.84$&$<3.0\times10^{-2}$&$<1.3\times10^{-2}$ \\
AB Aur    &  8.025 &oct04& 12.9 & $<1.2$ &$<449.1$&$<3.3\times10^{-2}$&$<3.9\times10^{-3}$ \\
          & 17.035 &dec02, jan04& 25.2 & $<1.1$ &$<4.5\times10^{-2}$&$<5.4\times10^{-3}$&$<4.0\times10^{-3}$ \\
AS 209    & 12.279 &jun03& 2.5 & $<0.61$ &$<4.2\times10^{-1}$&$<6.8\times10^{-3}$&$<3.1\times10^{-3}$ \\
AS 353    & 17.035 &jun03, jul03& 1.6 & $<1.6$ &$<7.1\times10^{-2}$&$<8.5\times10^{-3}$&$<6.4\times10^{-3}$ \\
FU Ori    & 12.279 &jan04, jan05& 6.4  & $<0.8$ &$<9.7$&$<1.6\times10^{-1}$&$<7.1\times10^{-2}$ \\
          & 17.035 &jan05& 5.8  & $<3.0$ &$<1.5$&$<1.8\times10^{-1}$&$<1.3\times10^{-1}$ \\
GG Tau    & 12.279 &jan04& 1.6  & $<2.1$ &$<2.0$&$<3.2\times10^{-2}$&$<1.5\times10^{-2}$ \\
          & 17.035 &nov01& 0.5  & $<2.8$ &$<1.1\times10^{-1}$&$<1.3\times10^{-2}$&$<9.7\times10^{-3}$ \\
GW Ori    & 12.279 &dec00, jan05& 8.6  & $<1.1$ &$<10.8$&$<1.7\times10^{-1}$&$<7.9\times10^{-2}$ \\
          & 17.035 &nov01& 4.9  & $<2.8$ &$<1.1$&$<1.3\times10^{-1}$&$<1.0\times10^{-1}$ \\
HD 163296 & 12.279 &jun03, jul03& 11.4 & $<0.6$ &$<4.3\times10^{-1}$&$<7.0\times10^{-3}$&$<3.2\times10^{-3}$ \\
          & 17.035 &jul03& 13.7 & $<2.35$ &$<6.9\times10^{-2}$&$<8.3\times10^{-3}$&$<6.2\times10^{-3}$ \\
IRAS 04278+2253 & 12.279&jan04, jan05 & 8.0 & $<0.62$ &$<5.9\times10^{-1}$&$<9.5\times10^{-3}$&$<4.3\times10^{-3}$ \\
                & 17.035 &jan05& 10.7 & $<2.8$ &$<1.1\times10^{-1}$&$<1.3\times10^{-2}$&$<9.7\times10^{-3}$ \\
L1551 IRS 5 & 12.279 &dec02& 13.2 & $<1.4$ &$<1.3$&$<2.1\times10^{-2}$&$<9.7\times10^{-3}$ \\
            & 17.035 &nov00& 20.0 & $<1.9$ &$<7.4\times10^{-2}$&$<8.8\times10^{-3}$&$<6.6\times10^{-3}$ \\
Lk H$\alpha$ 225 & 12.279 &jun03& 37.9 & $<2.5$ &$<121.7$&$<2.0$&$<8.9\times10^{-1}$ \\
                 & 17.035 &oct04& 48.2 & $<3.56$ &$<7.1$&$<8.4\times10^{-1}$&$<6.3\times10^{-1}$ \\
SVS 13 & 12.279 & dec02&13.3 & $<1.02$ &$<4.5$&$<7.2\times10^{-2}$&$<3.3\times10^{-2}$ \\
       & 17.035 & dec02&18.8 & $<2.0$ &$<3.6\times10^{-1}$&$<4.3\times10^{-2}$&$<3.2\times10^{-2}$ \\
V892 Tau & 12.279 &jan04& 31.2 & $<3.0$ &$<2.86$&$<4.6\times10^{-2}$&$<2.1\times10^{-2}$ \\
         & 17.035 &dec02& 71.7 & $<4.9$ &$<1.9\times10^{-1}$&$<2.2\times10^{-2}$&$<1.7\times10^{-2}$ \\
V1057 Cyg & 12.279 &oct04& 7.0 & $<0.8$ &$<14.0$&$<2.3\times10^{-1}$&$<1.0\times10^{-1}$ \\
Z CMa    & 12.279 &dec03& 142.6 & $<3.7$ &$<238.3$&$<3.8$&$<1.74$ \\
         & 17.035 &dec02& 170.8 & $<14.0$ &$<36.8$&$<4.4$&$<3.27$ \\ 
\enddata

\tablenotetext{a}{Measured TEXES flux value}
\tablenotetext{b}{3$\sigma$ limit in units of $10^{-14}$ ~ergs ~s$^{-1}$ ~cm$^{-2}$}   

\end{deluxetable}

\clearpage
\begin{deluxetable}{lrrrllll}
%preamble commands
\tabletypesize{\footnotesize}
%\rotate
\tablewidth{0pt}
%\tablenum{text}
\tablecolumns{8}
%\tableheadfrac{num}
\tablecaption{Summary of Upper Limits - Gemini/TEXES\label{tab:ulgemini}}
\tablehead{
&
&
&
&
&
&
&
\\
\colhead{Star} &
\colhead{$\lambda$} &
\colhead{Date} &
\colhead{F$_{\nu}$} &
&
\multicolumn{3}{c}{Mass in M$_{Jup}$} 
\\
\cline{6-8}
&
\colhead{($\mu$m)}&
&
\colhead{(Jy)} &
\colhead{Line Flux\tablenotemark{a,b}} &
\colhead{200 K} &
\colhead{500 K} &
\colhead{800 K} 
}

\startdata
%table data
49 Ceti & 12.279 & nov06&0.2\tablenotemark{c} & $<0.17$ &$<3.1\times10^{-2}$&$<5.0\times10^{-4}$&$<2.2\times10^{-4}$ \\
        & 17.035 & nov06&0.19\tablenotemark{c} & $<0.88$ &$<6.5\times10^{-3}$&$<7.7\times10^{-4}$&$<5.8\times10^{-4}$ \\
AS 209  & 12.279 & jul06&2.0\tablenotemark{d} & $<0.34$ &$<2.3\times10^{-1}$&$<3.8\times10^{-3}$&$<1.7\times10^{-3}$ \\
        & 17.035 & jul06&4.4\tablenotemark{e} & $<1.03$ &$<2.9\times10^{-2}$&$<3.4\times10^{-3}$&$<2.6\times10^{-3}$ \\
AS 353  & 12.279 & jul06&1.11\tablenotemark{f} & $<0.29$ &$<3.2\times10^{-1}$&$<5.1\times10^{-3}$&$<2.3\times10^{-3}$ \\
DoAr 21 & 12.279 & jul06&0.14\tablenotemark{g} & $<0.18$ &$<8.4\times10^{-1}$&$<1.4\times10^{-2}$&$<6.1\times10^{-3}$ \\
Elias 29& 17.035 & jul06&30.5\tablenotemark{e} & $<1.36$ &$<3.8\times10^{-2}$&$<4.6\times10^{-3}$&$<3.4\times10^{-3}$ \\
FU Ori  & 8.025  & nov06&3.5\tablenotemark{g}  & $<1.14$ &$<5143.6$&$<3.8\times10^{-1}$&$<4.5\times10^{-2}$ \\
        & 12.279 & nov06&3.6\tablenotemark{g}  & $<0.22$ &$<2.7$&$<4.3\times10^{-2}$&$<2.0\times10^{-2}$ \\
GM Aur  & 12.279 & nov06&0.25\tablenotemark{h} & $<0.54$ &$<5.2\times10^{-1}$&$<8.3\times10^{-3}$&$<3.8\times10^{-3}$ \\
GSS 30  & 17.035 & jul06&30.75\tablenotemark{f}& $<1.3$ &$<3.7\times10^{-2}$&$<4.3\times10^{-3}$&$< 3.3\times10^{-3}$\\
HD 141569& 12.279 & jul06&1.13\tablenotemark{e} & $<2.75$ &$<1.3$&$<2.1\times10^{-2}$&$<9.6\times10^{-3}$ \\
HD 163296& 12.279 & jul06&14.2\tablenotemark{e} & $<1.18$ &$<8.6\times10^{-1}$&$<1.4\times10^{-2}$&$<6.2\times10^{-3}$ \\
         & 17.035 & jul06&21.6\tablenotemark{e} & $<2.42$ &$<7.2\times10^{-2}$&$<8.5\times10^{-3}$&$<6.4\times10^{-3}$ \\
LkCa 15  & 12.279 & nov06&0.12\tablenotemark{g} & $<0.08$ &$<7.6\times10^{-2}$&$<1.2\times10^{-3}$&$<5.6\times10^{-4}$ \\
         & 17.035 & nov06&0.48\tablenotemark{e} & $<0.97$ &$<3.8\times10^{-2}$&$<4.5\times10^{-3}$&$<3.3\times10^{-3}$ \\
MWC 758  & 17.035 & nov06&3.8\tablenotemark{e} & $<0.65$ &$<5.2\times10^{-2}$&$<6.1\times10^{-3}$&$<4.6\times10^{-3}$ \\
RW Aur   & 8.025  & nov06&0.07\tablenotemark{f} & $<0.37$ &$<130.9$&$<9.7\times10^{-3}$&$<1.1\times10^{-3}$ \\
%         & 12.279 & 1.44\tablenotemark{f} & ??? & ??? & ??? & ???  \\
         & 17.035 & nov06&1.87\tablenotemark{f} & $<0.29$ &$<1.1\times10^{-2}$&$<3.7\times10^{-2}$&$<1.0\times10^{-3}$ \\
TW Hya   &12.279 & feb06& 0.5\tablenotemark{i} & $<0.6$ & $<7.6\times10^{-2}$ & $<1.2\times10^{-3}$ & $<5.5\times10^{-4}$\\
         &17.035 & feb06& 1.4\tablenotemark{i} & $<0.7$ & $<8.9\times10^{-2}$ & $<1.4\times10^{-3}$ & $<6.5\times10^{-4}$\\
V 1331 Cyg & 17.035 &nov06& 1.65\tablenotemark{f} & $<0.52$ &$<3.1\times10^{-1}$&$<3.7\times10^{-2}$&$<2.8\times10^{-3}$ \\
VV Ser   & 12.279 & jul06&4.61\tablenotemark{h} & $<0.55$ &$<1.6$&$<2.6\times10^{-2}$&$<1.2\times10^{-2}$ \\
         & 17.035 & jul06&2.6\tablenotemark{g} & $<1.2$ &$<1.4\times10^{-1}$&$<1.7\times10^{-2}$&$<1.3\times10^{-2}$ \\
\enddata

\tablenotetext{a}{3$\sigma$ limit in units of $10^{-14}$ ~ergs ~s$^{-1}$ ~cm$^{-2}$} 
\tablenotetext{b}{Upper limits calculated assuming FWHM=5.5 km s$^{-1}$ for TW Hya, 10 km s$^{-1}$ for others}   
\tablenotetext{c}{\citet{wahhaj07}}
\tablenotetext{d}{\citet{liu96}}
\tablenotetext{e}{ISO SWS archive}
\tablenotetext{f}{Spitzer IRS}
\tablenotetext{g}{Measured TEXES flux value}
\tablenotetext{h}{IRAS}
\tablenotetext{i}{\citet{ratzka07}}

\end{deluxetable}

%++++++++++++++++++++++++++++++++++++++++++++++++++++
% FIGURE CAPTIONS
%++++++++++++++++++++++++++++++++++++++++++++++++++++

\clearpage

\begin{figure}
\plotone{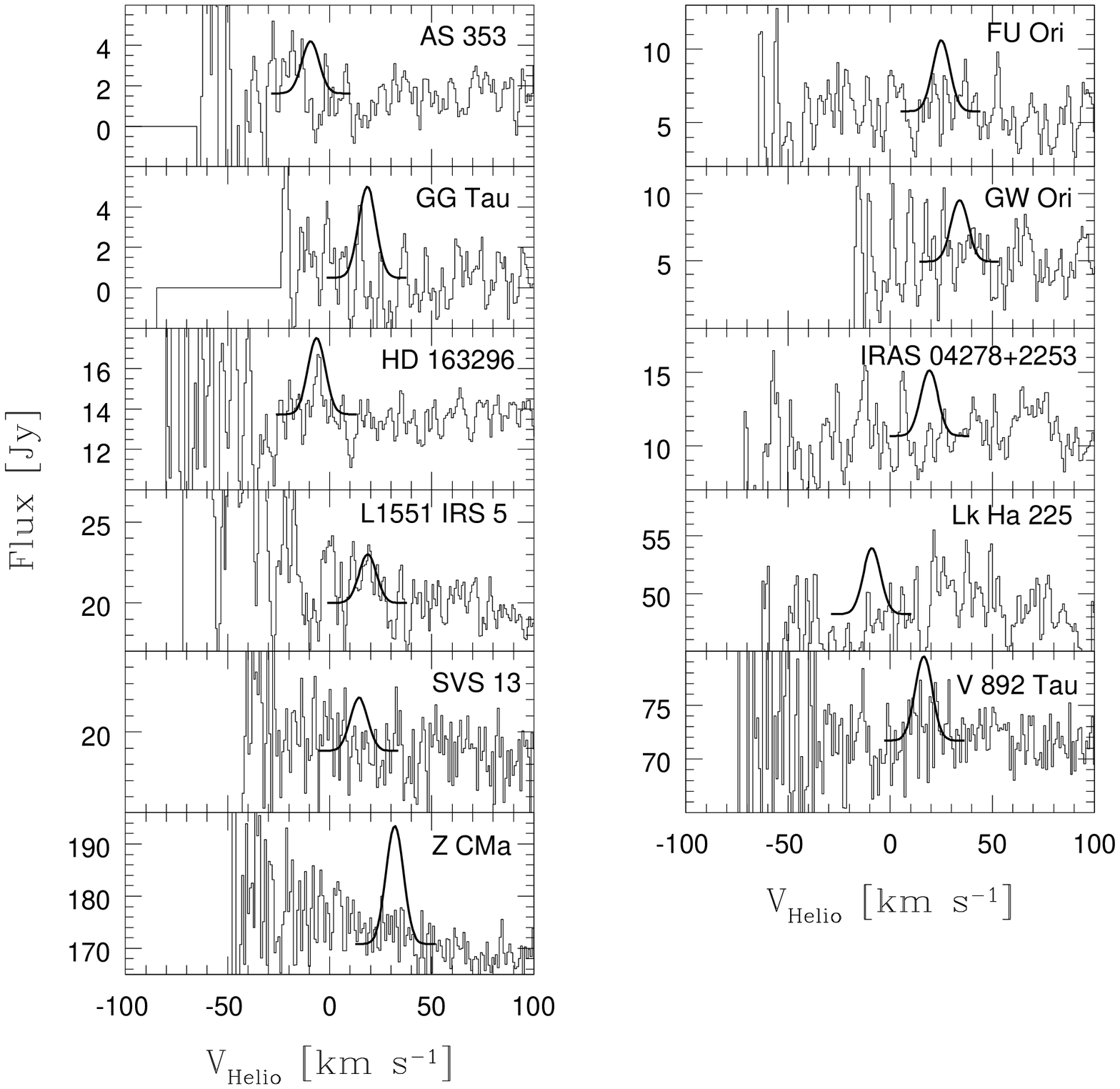}
%\centerline{f10.eps}
\figcaption[f10.eps]
{Upper limits for IRTF H$_{2}$ S(1) observations near 17 $\mu$m.  The overplotted 
Gaussian is centered at the stellar velocity of each source and represents 
the 3-$\sigma$ upper limit based on an assumed FWHM of 10 km s$^{-1}$. 
\label{irtf-s1}
}
\end{figure}
\clearpage

\begin{figure}
\plotone{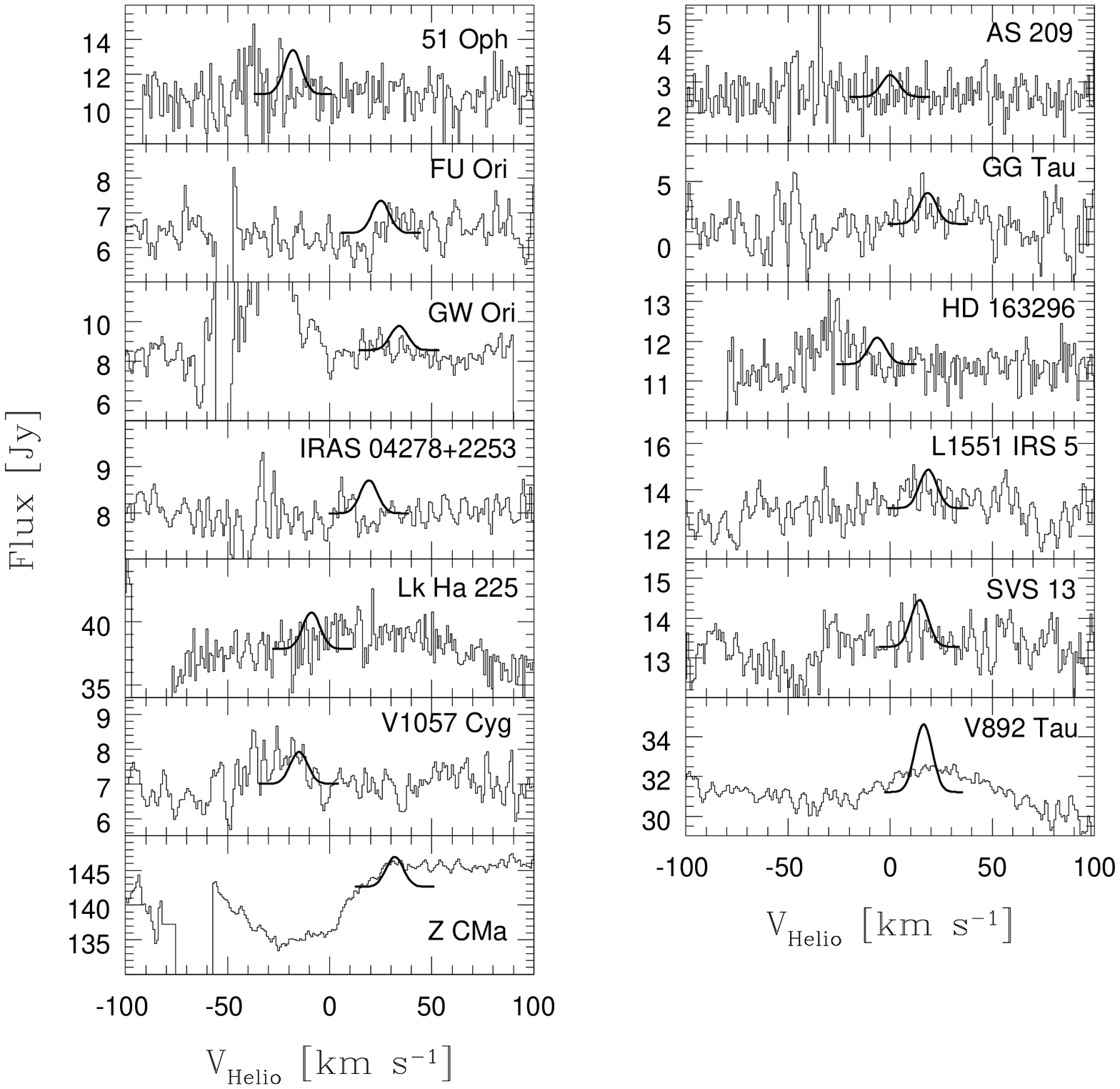}
%\centerline{f11.eps}
\figcaption[f11.eps]
{Upper limits for IRTF H$_{2}$ S(2) observations near 12 $\mu$m.  The overplotted Gaussian is 
centered at the stellar velocity of each source and represents the 3-$\sigma$ 
upper limit based on an assumed FWHM of 10 km s$^{-1}$.  
\label{irtf-s2}
}
\end{figure}
\clearpage

\begin{figure}
\plotone{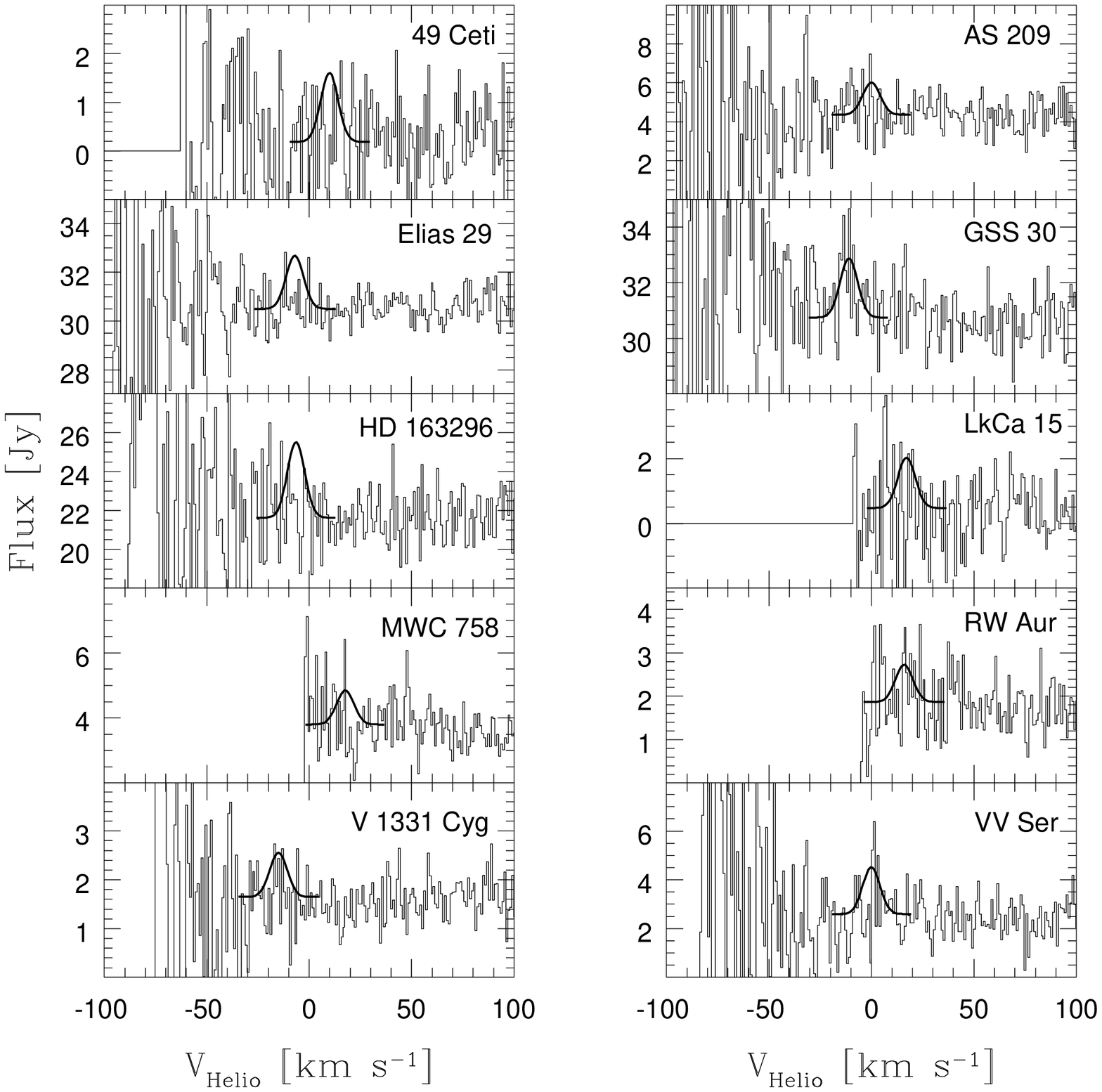}
%\centerline{f12.eps}
\figcaption[f12.eps]
{Upper limits for Gemini H$_{2}$ S(1) observations near 17 $\mu$m.  The overplotted Gaussian is 
centered at the stellar velocity of each source and represents the 3-$\sigma$ 
upper limit based on an assumed FWHM of 10 km s$^{-1}$. 
\label{gem-s1}
}
\end{figure}
\clearpage

\begin{figure}
\plotone{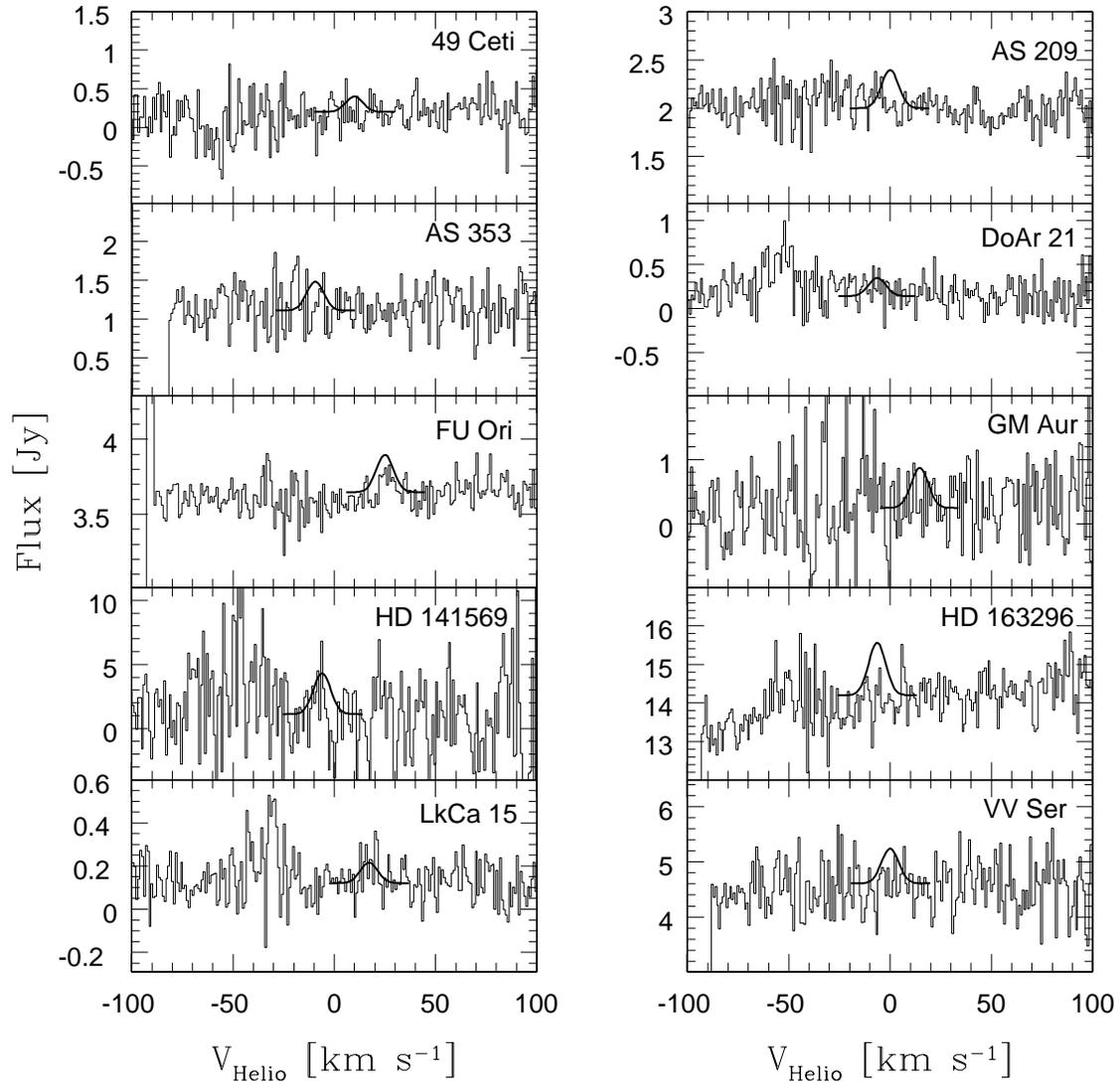}
%\centerline{f13.eps}
\figcaption[f13.eps]
{Upper limits for Gemini H$_{2}$ S(2) observations near 12 $\mu$m.  The overplotted 
Gaussian is centered at the stellar velocity of each source and represents the 3-$\sigma$ 
upper limit based on an assumed FWHM of 10 km s$^{-1}$. 
\label{gem-s2}
}
\end{figure}
\clearpage

\clearpage

\clearpage

%++++++++++++++++++++++++++++++++++++++++++++++++++++
%   APPENDIX
%++++++++++++++++++++++++++++++++++++++++++++++++++++

\clearpage

\appendix

\section{Accretion Shock Heating}

The total luminosity $L$ from a shock of area $A$ is given by:

\begin{equation}
L= {1 \over 2}\rho_o v_s^3 A,
\end{equation}
where $\rho _o$ is the preshock mass density and $v_s$ is the
shock velocity (the normal component of the velocity of
the flow with respect to the shock surface).  
The mass accretion rate $\dot M_{acc}$ through
the shock is given by:

\begin{equation}
\dot M_{acc} = \rho_o v_s A = m_p n_o v_s A,
\end{equation}
where $n_o$ is the gas hydrogen nucleus number density and $m_p$ is
the mass per hydrogen nucleus ($\sim 2.3 \times 10^{-24}$ gm).
Therefore, the luminosity can be rewritten

\begin{equation}
L= {1 \over 2} \dot M_{acc} v_s^2 .
\end{equation}
However, in an accretion shock onto an optically thick disk, where
one half of the radiation is emitted toward the disk midplane and
is absorbed, the escaping luminosity is given by \citep[see][]{neufeld94}:

\begin{equation}
L_{disk} = {1 \over 4} \dot M_{acc} v_s^2.
\end{equation}
The shock velocity is on the order of (but somewhat smaller due to the
oblique incident angle of the infall to the shock front) the freefall
velocity onto the disk, or, for our $r> 10$ AU constraint, $v_s \sim
5-10$ km s$^{-1}$. Assuming the accretion rate from the core onto
the disk through the accretion shock is similar to the accretion rate
from the disk onto the star, the measured accretion rates for our
sources are of order $\dot M_{acc} \sim 10^{-7}$ M$_\odot$ yr$^{-1}$.
Therefore,

\begin{equation}
L_{disk} \simeq 4 \times 10^{-4} \dot M_{-7} v_{s6}^2 \ \ {\rm L_\odot},
\end{equation}
where $\dot M_{-7} \equiv \dot M_{acc}/ 10^{-7}$ M$_\odot$ yr$^{-1}$
and $v_{s6} \equiv v_s/10^6$ cm s$^{-1}$= $v_s/10$ km s$^{-1}$.

To determine the luminosity in the pure rotational lines of H$_2$, one needs
the fraction $f_J$ of the total shock luminosity that emerges in the
H$_2$ 0-0 S(J) line.  This fraction depends on the preshock density $n_o$,
the shock velocity $v_s$, the amount of depletion of the preshock
gas coolants, and whether the shock is ``C type'' \citep{draine80} or
``J type'' \citep[cf.][]{hollenbach79}.  The preshock density can
be estimated by taking the shock area $A$ to be at least $2 \pi r_s^2
\sim 1.4 \times 10^{29}$ cm$^2$, with $r_s \sim 10$ AU and the factor
of 2 to account for both sides of the disk.  Using Eq. (2), we obtain

\begin{equation}
n_o \simeq 3 \times 10^6 \dot M_{-7} v_{s6}^{-1} A_{30}^{-1} \ {\rm cm^{-3}},
\end{equation}
with $A_{30} = A/10^{30}$ cm$^2$.  \citet{burton92} present results
for J shocks with $n_o = 10^6$ cm$^{-3}$ and $v_s = 5-10$ km s$^{-1}$ in
terms of the intensity $I_J$ of an H$_2$ 0-0 S(J) line.  Here,

\begin{equation}
f_J = {{4 \pi I_j} \over {0.5 m_p n_o v_s^3}}.
\end{equation}
They show cases with high abundances of gas phase oxygen not in CO and 
with extremely low abundances of gas phase oxygen not in CO.  The latter
case is perhaps more realistic since oxygen not in CO is expected to 
freeze out as water ice in dense cores, and a slow 5-10 km s$^{-1}$
shock does not release water from the ice mantles to the gas \citep{hollenbach79}. 
In the former case, the gas phase oxygen not in CO rapidly
converts to gas phase H$_2$O in the shock, and the H$_2$O dominates
the shock cooling and thereby weakens the H$_2$ lines.  \citet[see Figures 5b and 6]{burton92} 
find that for the case with H$_2$O 
freezeout $f_1 \sim 2.5 \times 10^{-3}$, $f_2 \sim 2.5 \times 10^{-3}$
and $f_3 \sim 10^{-2}$. In the case with abundant gas phase oxygen not
in CO, $f_1 \sim 10^{-4}$, $f_2 \sim 10^{-4}$, and $f_3 \sim 10^{-3}$.
\citet{burton92} do not present results for $f_4$, but inspection of the
unpublished output of these runs reveals $f_4 \sim 0.5 f_3$.  A lower
$v_s$ results in lower ratios of $f_4/f_2$.

\citet{draine83} present $I_J$ for C type shocks with
$n_o= 10^6$ cm$^{-3}$, preshock magnetic field component parallel to
the shock front $B_o = 0.5$ mG, electron abundance $x_e = 10^{-8}$, and
where gas phase oxygen not in CO is {\it not} depleted.  Their result
at $v_s= 10$ km s$^{-1}$ implies $f_1 \sim 10^{-2}$, $f_2 \sim 4 \times
10^{-3}$, and $f_3 \sim 10^{-2}$.  They also do not present results
for $f_4$, but we again estimate $f_4 \sim 0.5 f_3$.  Presumably,
these fractions would be somewhat larger if oxygen is allowed to 
freezeout as water ice.

From the above results, we see for J shocks with water ice freezeout
and for C shocks, the luminosities in the H$-2$ 0-0 S(1), S(2) and S(4)
lines are $\sim 10^{-6} - 10^{-5} \dot M_{-7} v_{s6}^2$ L$_\odot$.
These luminosities (assuming $\dot M_{-7} \sim 1$ correspond to our
observed luminosities, suggesting accretion shocks as a viable excitation mechanism
for these emission lines.  For J shocks without freezeout the luminosities
are about an order of magnitude less than observed for $\dot M_{-7} \sim 1$.

These same shock results can be used to estimate the intensities of CO
mid J transitions in the shocks.  \citet{burton92} show that in the
J shocks presented here the CO J= 17-16 transition is $\sim 3$ times
stronger than the H$_2$ 0-0 S(1) line.  \citet{draine84} and \citet{draine83} 
present C shock models
from which the CO J$\rightarrow$ J-1 line strengths can be estimated.
For $n_o = 10^6$ cm$^{-3}$, $v_s = 10$ km s$^{-1}$, and $B_o$ and $x_e$ from
above, the CO J $\rightarrow$ J-1 line intensities peak at J $\sim 10 -
12$ with strengths comparable to that of H$_2$ 0-0 S(1).  Thus, Herschel
observations of these disks with the HIFI instrument may detect the 
submillimeter CO lines near the J peak, validate the shock models,
and constrain the shock parameters such as $n_o$, $v_s$, and $A$.

It is difficult to estimate the strengths of lines other than those
of H$_2$ and CO because their strength depends on their gas phase
abundances, and all species other than the undepleted H$_2$ and
CO (e.g., Fe, Fe$^+$, S, O) may be heavily depleted on grains as
ice mantles and/or refractory material.  Such slow shocks are 
unlikely to remove them.  Radiative transfer calculations need to
be performed to see if the dust grains are warm enough to
thermally sublimate the ice mantles at distances of 10-30 AU
from the star.  A more detailed shock modeling of some
of these observed sources is warranted.

\end{document}